\documentclass[twocolumn,aps,pra,superscriptaddress]{revtex4-1}
\usepackage[latin9]{inputenc}
\usepackage{amsmath}
\usepackage{amssymb}
\usepackage{graphicx}
\usepackage{array}
\usepackage{times}
\usepackage[usenames,dvipsnames]{color}
\makeatletter
\usepackage{dcolumn}
\usepackage{bm}

\makeatother

\begin{document}

\title{Universal stabilization of single-qubit states using a tunable coupler}
\author{Ziwen Huang}
\thanks{These authors contributed equally to this work.}
\affiliation{Department of Physics and Astronomy, Northwestern University, Evanston, Illinois 60208, USA}
\author{Yao Lu}
\thanks{These authors contributed equally to this work.}
\affiliation{James Franck Institute and Department of Physics, University of Chicago,
Chicago, Illinois 60637, USA}
\author{Eliot Kapit}
\affiliation{Department of Physics and Engineering Physics, Tulane University, New Orleans, Louisiana 70118, USA}
\author{David I.\ Schuster}
\affiliation{James Franck Institute and Department of Physics, University of Chicago,
Chicago, Illinois 60637, USA}
\author{Jens Koch}
\affiliation{Department of Physics and Astronomy, Northwestern University, Evanston, Illinois 60208, USA}

\date{\today }
\begin{abstract}
\small
We theoretically analyze a scheme for fast stabilization of arbitrary qubit states with high fidelities, extending a protocol recently demonstrated experimentally  [Lu \textit{et al.}, Phys.\,Rev.\,Lett.\,119,\,150502\,(2017)]. That experiment utilized red and blue sideband transitions in a system composed of a fluxonium qubit, a low-$Q$ LC-oscillator, and a coupler enabling us to tune the interaction between them. Under parametric modulations of the coupling strength, the qubit can be steered into any desired pure or mixed single-qubit state. For realistic circuit parameters, we predict that stabilization can be achieved within $100\,$ns.
By varying the ratio between the oscillator's damping rate and the effective qubit-oscillator coupling strength, we can switch between under-damped, critically-damped, and over-damped stabilization and find optimal working points. We further analyze the effect of thermal fluctuations and show that the stabilization scheme remains robust for realistic temperatures.
\end{abstract} 
\maketitle

\section{Introduction}
Superconducting quantum circuits are among the most promising platforms for quantum computing, offering great flexibility and potential for scalability by microfabrication techniques \cite{IBM1,JQYou,AAHouck,DevoretScience,3Dqubit,3Dqubit2,3Dqubit3,IBMgate,NoriReview,Liu_sc_ion}. Strategies for stabilizing desired qubit states on demand constitute important building blocks for future error-tolerant circuit QED networks, fulfilling tasks such as qubit reset, initialization and entanglement generation \cite{Peropadre,PCIActive,VijayNature,GeerlingReset,Leghtas,EliotPra,EliotPRL,Barbara,Cohen,ShankarNature}. In the past, several schemes have been explored for stabilizing single-qubit \cite{Murch,GeerlingReset,VijayNature,PCIActive} and multi-qubit states \cite{Leghtas, ShankarNature,EliotPra,EliotPRL}, using active feedback \cite{PCIActive,VijayNature} or autonomous stabilization \cite{GeerlingReset,Leghtas,EliotPra,EliotPRL,Murch,ShankarNature}. The latter schemes employ engineered dissipation processes \cite{Eliot2017} to counteract undesired decoherence and protect specific quantum states.  Murch et al.\ \cite{Murch} have demonstrated a pioneering scheme that can stabilize arbitrary single-qubit states, which is an important step towards implementing error-correction code.

Over the last decade, tunable-coupler devices in quantum circuit networks have yielded a variety of achievements \cite{Yao,Swflux,Liu_fluxq,ABtunable,Vanderploeg,IBMTunable,Bertet,AllmanTunable,Peropadre,YuchenTunable,Nori_Interqubit}. Researchers have shown that parametric modulations in tunable-coupler circuits can generate flexible photon-conserving and non-conserving qubit-qubit and qubit-resonator couplings in rotating frames \cite{AllmanTunable,Peropadre,Bertet,Leeksideband,Wallraffsideband,Eliot2015,Yao}. These induced interactions, often referred to as red- and blue-sideband couplings, are tunable and can also serve as useful resources for implementing qubit state stabilization \cite{Yao, Leeksideband}. In recent work \cite{Yao}, we have experimentally demonstrated how engineered dissipation and tunable coupling may be combined to realize universal qubit stabilization. In the tunable-coupling architecture of this experiment,  a transmon qubit and a low-$Q$ resonator are coupled by a dc superconducting quantum interference device (SQUID) loop. Red and blue-sideband interactions between  resonator and qubit can then be produced by modulating the magnetic flux penetrating the SQUID loop.

Based on our previous work, here we present a different full universal stabilization protocol, which can access both pure and mixed qubit states. Key to achieving this is the joint use of two flux modulation tones and a Rabi drive tone. A large qubit anharmonicity is desirable for this scheme to work, therefore we choose a fluxonium qubit in our circuit, instead of a transmon qubit used in the previous paper \cite{Yao}. 
Our analysis shows that optimization should allow for stabilization fidelities of over $99.5\%$ for any pure qubit state with realistic circuit parameters and operation temperatures. We analytically derive the stabilization times and critical damping parameters based on the Lindblad master equation. In contrast to previous fixed-coupling schemes \cite{Murch,GeerlingReset,HollandFock}, we do not require large photon numbers, and the stabilization process can be completed within relatively short times of the order of 100$\,$ns for all qubit states. We can further achieve stabilization of mixed states and tune the purity of the stabilized state via the coupling-strength ratio. In this sense, any single-qubit state on and inside the Bloch sphere can be targeted by this scheme.

The outline of our paper is as follows. In Section II, we show the derivation of the Hamiltonian with red and blue-sideband couplings, closely following the idea of the quantum circuit realized in Ref.~\cite{Yao}. Section III details the single-qubit stabilization scheme, starting from $z$-axis stabilization, and then generalizing to arbitrary-axis stabilization. We systematically study the dependence of the stabilization fidelity on dissipation rates, driving strengths and temperature, providing both analytical approximations of the fidelity as well as results from numerical simulations. We  investigate the stabilization dynamics and analyze the stabilization time of the pure-state stabilization process in Section IV, and finally provide our conclusions in Section V.

\section{Model of tunable circuit}
\begin{figure}
\includegraphics[width=8.5cm]{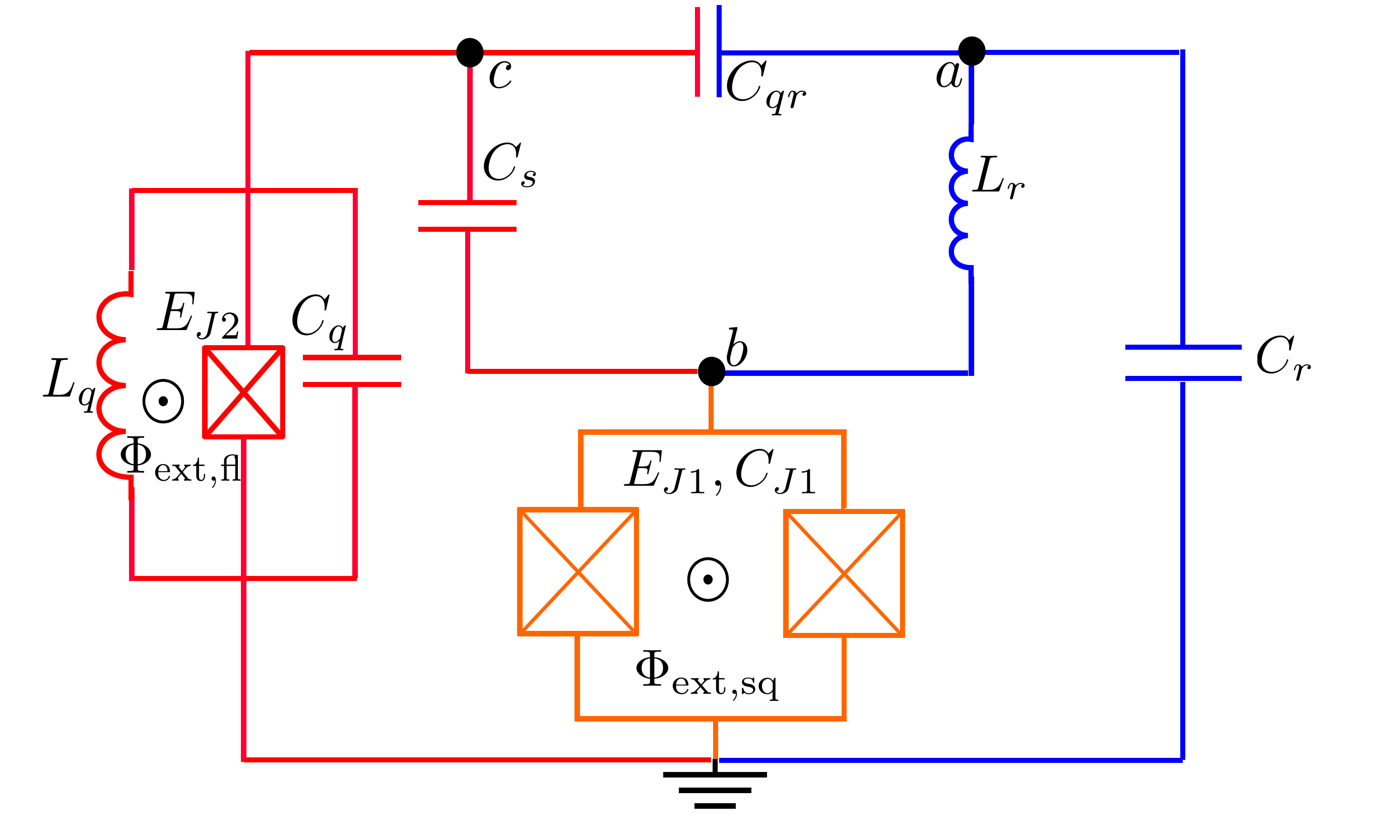}
  \caption{ Circuit diagram of the device. The three nodes $a$, 
  $b$, and $c$ belong to resonator mode, coupler mode and qubit mode, respectively. The coupler consists of a dc SQUID, which provides the an effective Josephson energy tunable via the magnetic flux $\Phi_{\mathrm{ext,sq}}$. This, in turn, alters the effective coupling strength between the qubit and resonator.}\label{fig0}
\end{figure}

Our stabilization protocol is based on the device shown in Fig.~{\ref{fig0}}. The superconducting circuit consists of three components: a lumped-element resonator, a fluxonium qubit, both connected in parallel with a dc SQUID, serving as an effective coupler. The coupler is similar to a tunable inductor shared between the resonator and qubit, the inductance of which can be tuned by external flux.

Circuit analysis and quantization (see Appendix A for details) yield the following Hamiltonian for the circuit:
\begin{align}
\label{FullH}
H=&\,4E_an_a^2+4E_bn_b^2+4E_cn_c^2+E_{ac}n_an_c+E_{bc}n_bn_c\nonumber\\
&-E_{J1\mathrm{eff}}(t)\cos\varphi_b-E_{J2}\cos(\varphi_c+\varphi_{\mathrm{fl}})\nonumber\\
&+\frac{E_{Lr}}{2}(\varphi_a-\varphi_b)^2+\frac{E_{Lq}}{2}\varphi^2_c.
\end{align}Here, the three nodes $i=a$, $b$, and $c$ belong to the resonator, coupler and qubit degrees of freedom.
The conjugate variables $n_i$ and $\varphi_i$ denote the corresponding charge and phase operators, and $E_i$ the associated charging energies.   $E_{ac}$ and $E_{bc}$ are the capacitive coupling energies, $E_{Lr,q}=(\Phi_0/2\pi)^2/L_{r,q}$ stand for inductive energies of the resonator and qubit, and $E_{J1,2}$ are the Josephson energies of the coupler and qubit junctions, respectively.  We denote the external magnetic fluxes penetrating the fluxonium and dc-SQUID loops by $\Phi_{\mathrm{ext,sq}}$ and $\Phi_{\mathrm{ext,fl}}$, while  $\varphi_{\mathrm{sq}}=2\pi\Phi_{\mathrm{ext,sq}}/\Phi_0$ and $\varphi_{\mathrm{fl}}=2\pi\Phi_{\mathrm{ext,fl}}/\Phi_0$ represent the corresponding reduced fluxes. The former tunes the effective Josephson energy of the coupler following the relation $E_{J1\mathrm{eff}}(t)=2E_{J1}\cos[\varphi_{\mathrm{sq}}(t)/2]$, and the latter is slightly modulated around zero flux for the generation of a Rabi drive (details are presented in Appendix A). 

By design, the coupler mode has an excitation energy far exceeding those of the qubit and resonator, and exclusively fulfills the passive function of mediating the coupling between the resonator and qubit. The effective Hamiltonian, reduced to resonator and qubit modes only, is obtained by adiabatically eliminating the coupling terms among the three modes. Specifically, we perform a Bogoliubov and a Schrieffer-Wolff transformation to decouple the three modes and integrate out the coupler mode (see details in Appendix A), assuming that the coupler mode remains in its ground state throughout. Dynamic modulation of the external magnetic flux threading the SQUID loop then leads to an effective tunable coupling between the dressed resonator and qubit modes, whose strength we denote by $g(t)$ (see derivation details in Appendix A). The resulting effective Hamiltonian, in the dressed basis, is given by ($\hbar = 1$)
\begin{align}
\label{FullH2}
H'=&\omega_r a^\dagger a+\omega_q\sigma^+\sigma^--\chi\sigma_z a^\dagger a\nonumber\\
&+g(t)(a^\dagger+a)(\sigma^++\sigma^-).
\end{align}Here, $\omega_r/2\pi$ and $\omega_q/2\pi$ are the dressed resonator and qubit frequencies, and $
\chi$ stands for the dispersive shift. (Since the expressions for the dressed frequencies and the dispersive shift are lengthy, they are relegated to Appendix A). Dynamic modulation of $\Phi_{\mathrm{ext,sq}}$ at different frequencies can generate sideband interactions \cite{Wallraffsideband} between the resonator and qubit modes. For our stabilization scheme, we modulate the flux with two tones of frequencies $\omega_1$ and $\omega_2$. The time-dependent coupling $g(t)$ generated by this can be approximated by
\begin{align}
g(t)\approx g(2\epsilon_1\cos\omega_1t+2\epsilon_2\cos\omega_2t),
\label{gt}
\end{align} 
where $\epsilon_{1,2}$ parametrizes the amplitudes of the modulation tones. 

In the rotating frame reached by the unitary transformation
\begin{align}
\label{Unitary}
U=\exp[i\omega_r ta^\dagger a +i\omega_q t\sigma^+\sigma^-],
\end{align}
the effective Hamiltonian takes on the form
\begin{align}
\label{FullH3}
\tilde{H}\approx &\, g(2\epsilon_1\cos \omega_1 t+2\epsilon_2 \cos\omega_2 t)\nonumber\\
&\times(a^\dagger \mathrm{e}^{i\omega_r t}+\text{H.c.})(\sigma^+\mathrm{e}^{i\omega_q t}+\text{H.c.})-\chi\sigma_za^\dagger a.
\end{align}
Then, with modulation frequencies matching the difference and the sum of resonator and qubit frequencies \cite{Wallraffsideband}, $\omega_1 = \omega_r-\omega_q$ and $\omega_2 =  \omega_r+\omega_q$, we arrive at the Hamiltonian essential for the implementation of our stabilization scheme,
\begin{align}
\label{Hrb}
\tilde{H}\approx g\epsilon_1(a^\dagger \sigma^-+a\sigma^+)+g\epsilon_2(a^\dagger\sigma^++a\sigma^-)-\chi\sigma_z a^\dagger a.
\end{align}
The ac-Stark shift term is a remnant not helping our stabilization scheme and should thus be made  small. In the following discussion, we will neglect this term and then validate our approximation numerically.

\section{Qubit stabilization}
In this section we describe the single-qubit stabilization scheme, and discuss the dependence of stabilization fidelity on drive strength, dissipation rates and temperature. We show that we can stabilize the qubit in any pure state on the Bloch sphere as well as in any mixed state, along any desired stabilization axis. The following discussion assumes a sufficiently large qubit anharmonicity, such that the qubit can simply be modeled as a two-level system. We start our discussion with stabilization of the  qubit along the $z$-axis, and then generalize to arbitrary axis.


This protocol differs from existing approaches in a few key ways. In previous proposals \cite{Yao,Murch}, a detuned ac drive is applied to the qubit, generating a uniform magnetic field Hamiltonian for the qubit's pseudospin in the rotating frame. The direction and magnitude of this field are determined from the phase, Rabi frequency, and detuning of the ac drive. The qubit is then coupled to a lossy resonator through a coupler with fixed direction on the Bloch sphere.  The drive frequencies are chosen such that a particular state is stabilized by an energetic resonance condition, set by the splitting between the two qubit states in the rotating-frame Hamiltonian. Because of this, the maximum fidelity is limited by the size of that splitting, which is typically small. In contrast, our protocol leaves the qubit alone and varies both the magnitude and direction of the tunable coupling, ensuring that a particular state is chosen for stabilization at the operator level rather than through energy matching. This will allow for substantially higher maximum fidelities, as we will now show.

\subsection{Stabilizing the qubit along the $z-$axis}
\begin{figure}
\includegraphics[width=6cm]{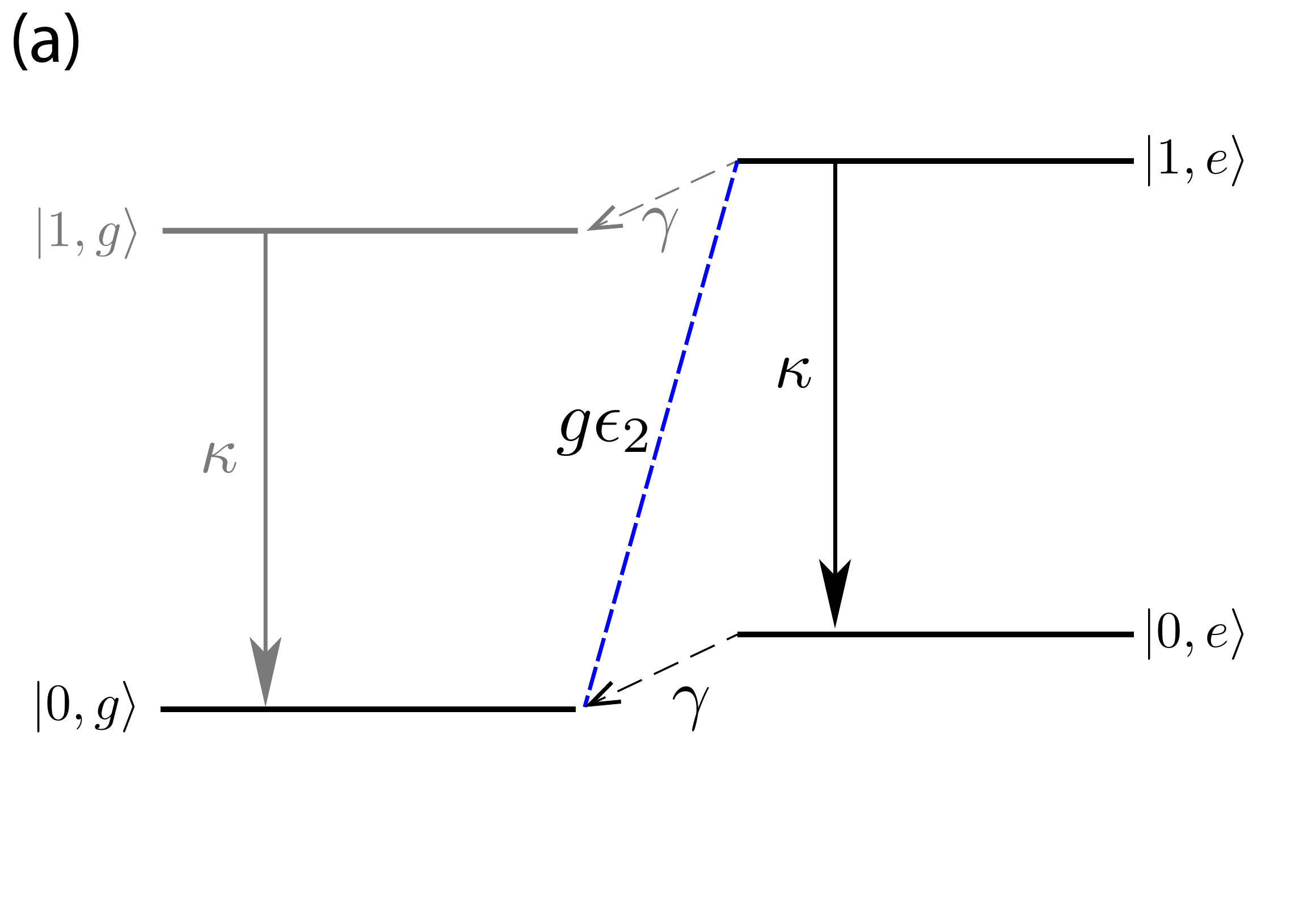}
\includegraphics[width=6cm]{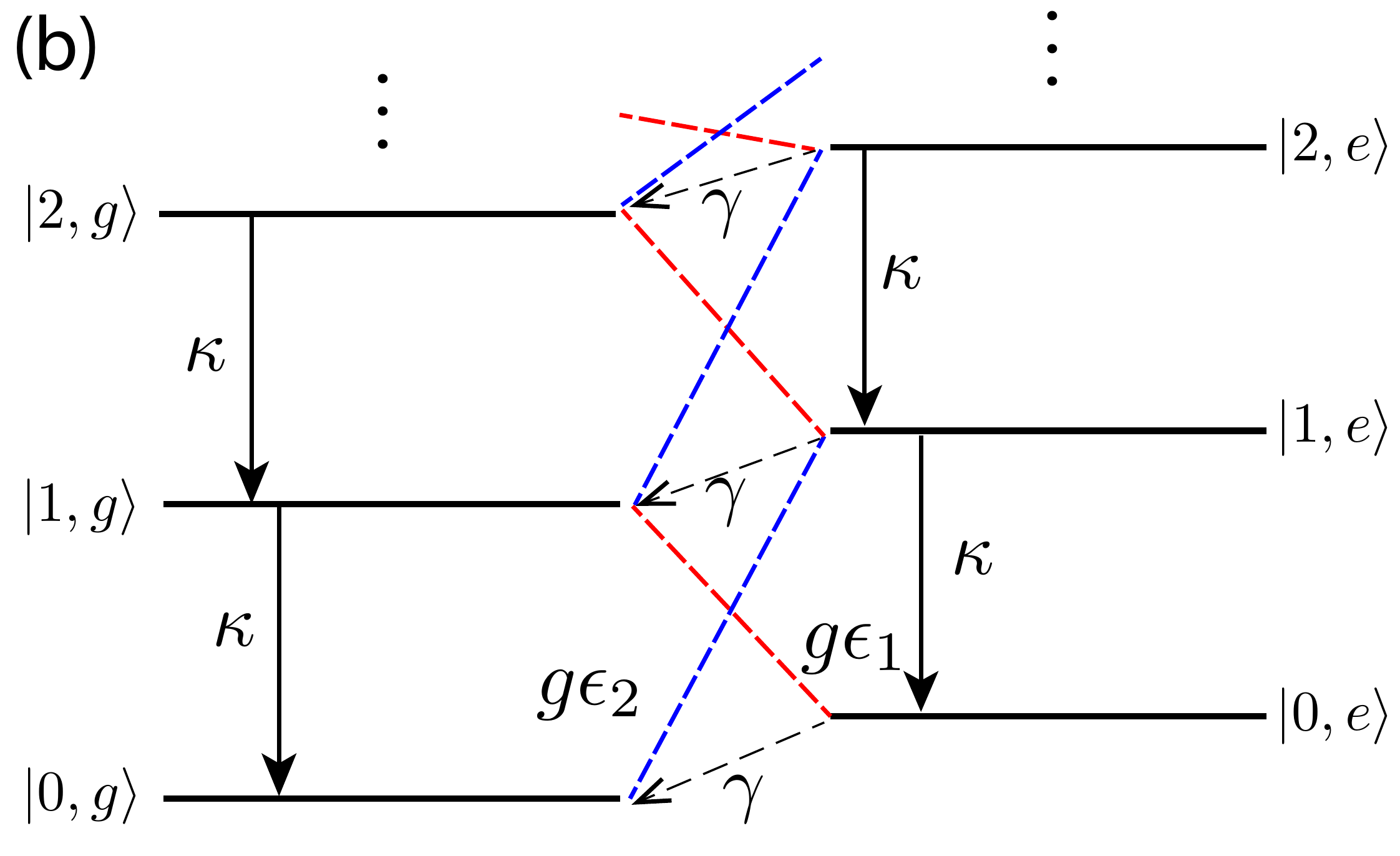}
\caption{(a) shows the ladder diagram with only blue-sideband coupling turned on; (b) shows the diagram with both red and blue-sideband couplings. In (b), for $\kappa\gg\gamma$, population in $\vert 1,g\rangle$ can be safely neglected}\label{fig1}
\centering
\end{figure}

Stabilizing the qubit's excited state in the presence of relaxation is done via blue-sideband coupling and fast resonator decay. This idea was first proposed in Ref.~\cite{Leeksideband}, and has been implemented in experiments \cite{Leeksideband,Yao}. For blue sideband only, we merely need the to modulate the flux at the sum frequency ($\epsilon_1=0$) and thus obtain the effective Hamiltonian 
\begin{align}
H_b=g\epsilon_2(a^\dagger\sigma^++a\sigma^-).
\label{Hb}
\end{align} 
Here, blue-sideband coupling strength $g\epsilon_2$ and resonator dissipation rate $\kappa$ should be chosen much larger than the qubit decay  and dephasing rates $\gamma$, $\Gamma_{\varphi}$. The stabilization mechanism is highlighted in Fig.~\ref{fig1}(a), showing the relevant energy eigenstates and processes leading to coherent and incoherent transitions among them. The blue-sideband terms (blue dashed line) couple the states $\vert m+1,e\rangle$ and $\vert m,g\rangle$, where $m$ stands for the photon occupation number in the resonator, and $g$ and $e$ denote the qubit ground and excited states. Qubit relaxation and photon decay are marked by arrows. To assess the dynamics of the system, consider a quantum trajectory starting in the ground state $\vert 0,g\rangle$. The blue-sideband coupling quickly shifts occupation amplitude to the state $\vert 1,e\rangle$ on the time scale $\sim(g\epsilon_2)^{-1}$. The $\vert 1,e\rangle$ state will typically lose its photon in a short time $\sim \kappa^{-1}$ and thus enter the target state $\vert 0,e\rangle$. Relative to the time scales involved so far, qubit decay is slow. Whenever the qubit does induce the system to return to the ground state $\vert 0,g\rangle$, the described process starts over, thus making $\vert 0,e\rangle$ the state predominantly occupied during the dynamics. In other words, the system will be stabilized in $\vert 0,e\rangle$ with $\langle a^\dagger a\rangle\approx 0$ and $\langle\sigma_z\rangle\approx 1$.

For our analytic treatment, we neglect population in $\vert 1,g\rangle$, since quick photon decay is expected to prevent occupation amplitude to build up in this state.
We thus consider the dynamics of the system in the subspace spanned by $\vert 0,g\rangle$, $\vert 0,e\rangle$ and $\vert 1,e\rangle$. The evolution of the system, at zero temperature, is governed by the Lindblad master equation,
\begin{align}
\frac{\mathrm{d}\rho}{\mathrm{d}t}=-i[H_b,\rho]+\kappa\,\mathbb{D}[a]\rho+\gamma\,\mathbb{D}[\sigma^-]\rho+\frac{\Gamma_\varphi}{2}\,\mathbb{D}[\sigma_z]\rho,
\end{align} 
where we truncate the density matrix $\rho$ and all other operators to the three levels of relevance. In the equation above,  the damping superoperator is defined by $\mathbb{D}[L]\rho=(2L\rho L^\dagger-L^\dagger L\rho-\rho L^\dagger L)/2$. We assess the stabilization performance by calculating the state fidelity for the qubit's excited state, $\mathcal{F}_z=\sqrt{\langle e\vert \rho_q\vert e\rangle}$, where $\rho_q$ is the qubit's reduced density matrix. By solving for the steady state, ${\mathrm{d}\rho}/{\mathrm{d}t}=0$, we  obtain an analytical expression for this stabilization fidelity:
\begin{align}
\label{sz}
\mathcal{F}_z=\sqrt{1-\left[\frac{2g\epsilon_2}{\kappa}+\left(\frac{1}{2}\kappa+\Gamma_\varphi\right)\frac{1}{g\epsilon_2}\right]C},
\end{align}
where 
\begin{align}
\label{C}
C=&\,\mathrm{Im}[\langle 0,g\vert \rho\vert 1,e\rangle]\nonumber\\
=&\left[\frac{2g\epsilon_2}{\gamma}+\frac{2g\epsilon_2}{\kappa}+\left(\frac{1}{2}\kappa+\Gamma_\varphi\right)\frac{1}{g\epsilon_2}\right]^{-1}.
\end{align}
In the limit $2g\epsilon_2/\gamma\gg 2g\epsilon_2/\kappa$, $(\kappa/2+\Gamma_\varphi)/g\epsilon_2$, one obtains the more compact approximation
\begin{align}
\label{appr}
\mathcal{F}_z \approx \sqrt{1-\left[\frac{2g\epsilon_2}{\kappa}+\frac{\kappa}{2g\epsilon_2}\right]\frac{\gamma}{2g\epsilon_2}}.
\end{align}
For given qubit dissipation rates, we can optimize the state fidelity by tuning the resonator dissipation rate $\kappa$ and modulation strength $g\epsilon_2$. First, considering fixed $\kappa$, the fidelity increases monotonically with $g\epsilon_2$ and approaches an upper limit set by $\lim_{g\epsilon_2\to\infty}\mathcal{F}_z = \sqrt{1-\gamma/\kappa}$. For fixed $g\epsilon_2$, Eq.~(\ref{appr}) shows that the fidelity approximately reaches its maximum for $\kappa=2g\epsilon_2$,
namely
\begin{align}
\max_{\kappa>0}\mathcal{F}_z\approx \sqrt{1-\frac{\gamma}{g\epsilon_2}}.
\label{optimal2}
\end{align}
Fig.~{\ref{fig2}(a) shows numerical results for the fidelity as a function of $g\epsilon_2$ and $\kappa$, obtained by a full simulation of the steady state based on Eq.~(\ref{Hb}). We find that high stabilization fidelities exceeding 99.5\% can be reached with realistic parameters. 
The optimum condition  $\kappa=2g\epsilon_2$  is shown as the dashed line on the $\kappa$-$g\epsilon_2$ plane, which yields the maximum fidelity values.
\begin{figure}
\centering
\includegraphics[width=8cm]{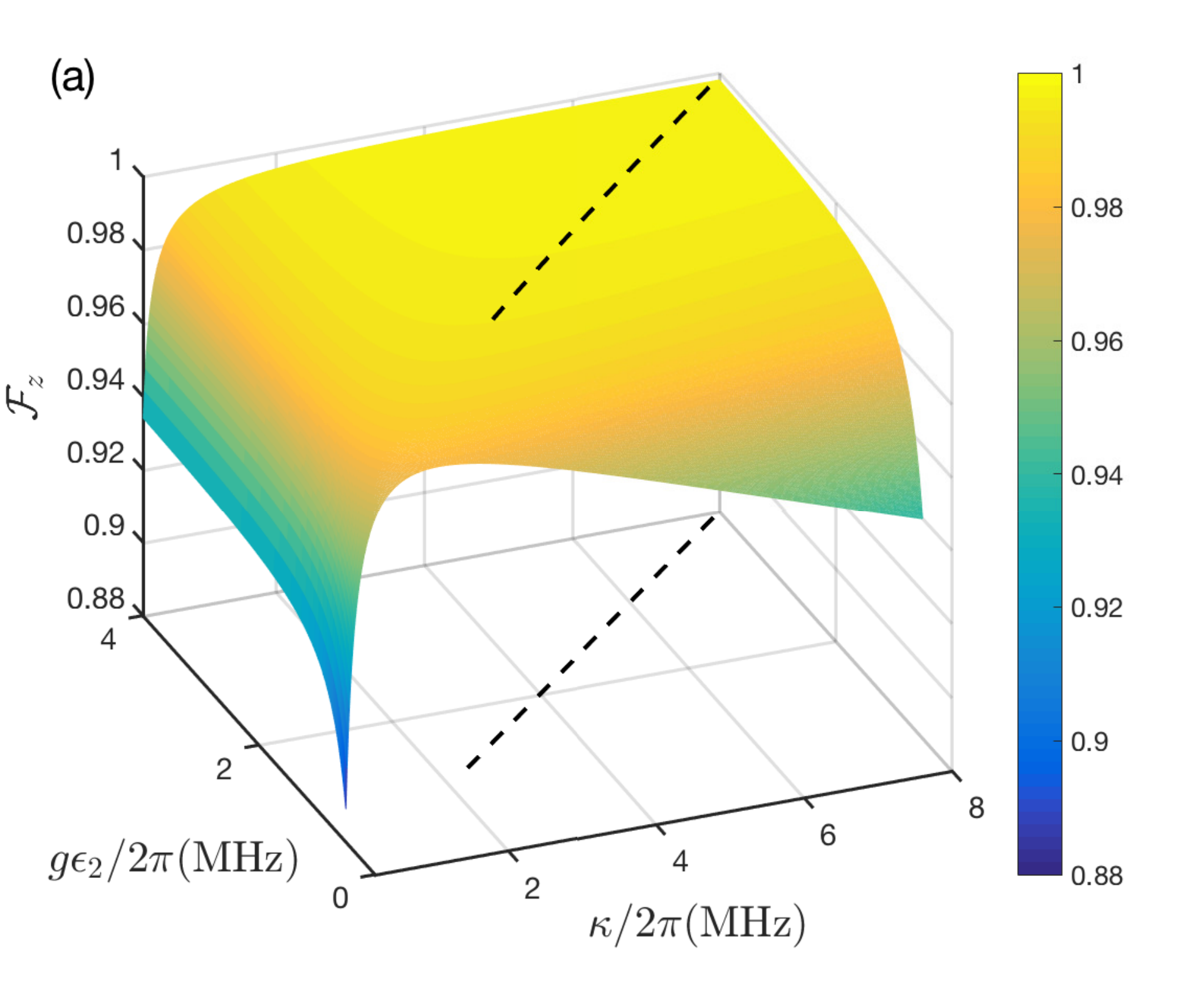}
\includegraphics[width=6.4cm]{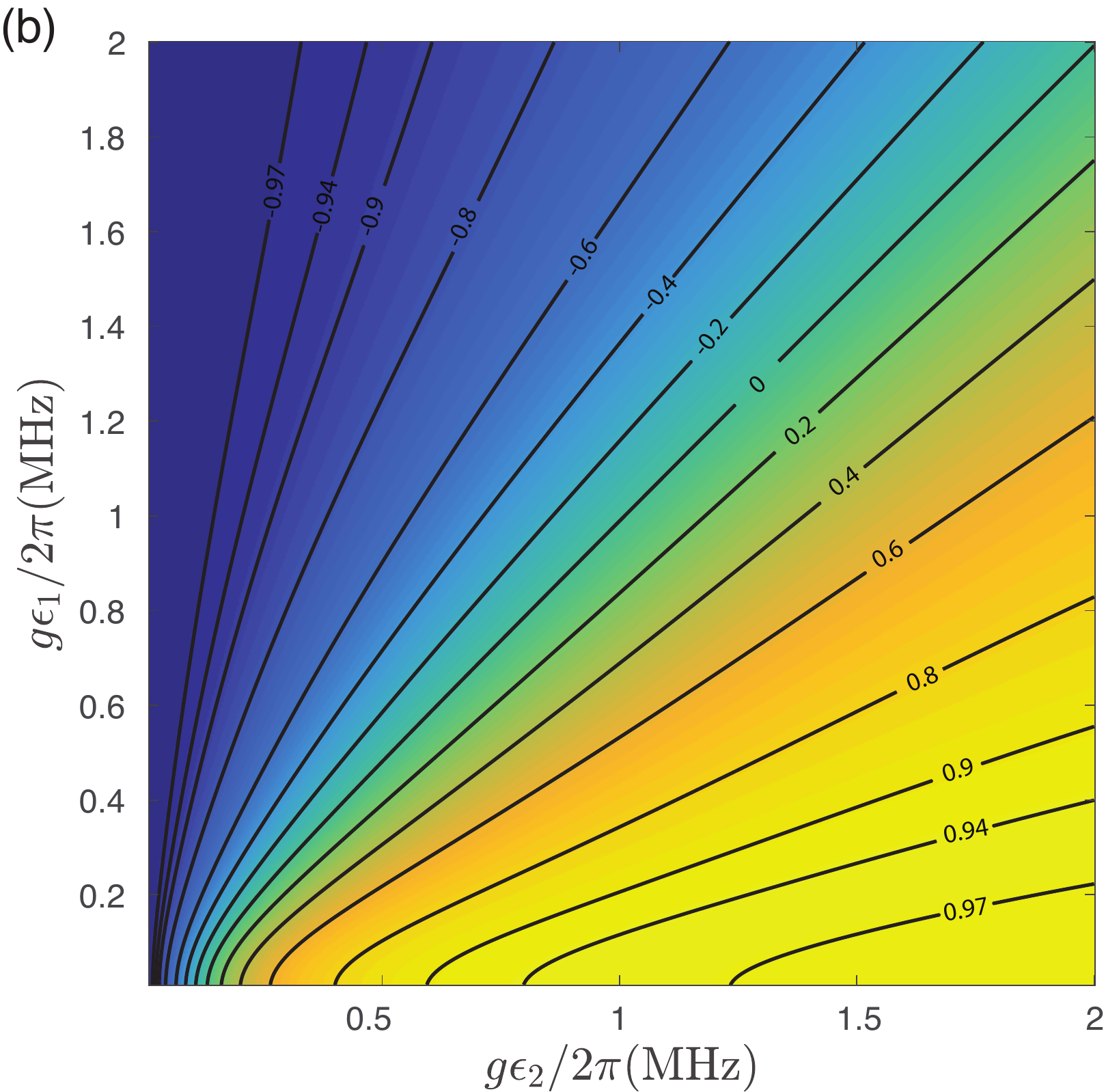}
\caption{(a) Fidelity for stabilization in the excited state $\vert e\rangle$, as a function of resonator dissipation rate $\kappa$ and blue-sideband coupling strength $g\epsilon_2$. The dotted line is the approximated maximum line from Eq.~(\ref{appr}). (Qubit dissipation rates are chosen as $\gamma=\Gamma_{\varphi}=0.1\,$MHz.) (b) Expectation of $\sigma_z$ for different $g\epsilon_1$ and $g\epsilon_2$, see Eq.~(\ref{Hrb}), with a fixed $\kappa/2\pi=4\,$MHz. All results shown assume $\chi/2\pi=0.5\,\mathrm{MHz}$ and a temperature of 15$\,$mK.}
\label{fig2}
\centering
\end{figure}

 We note from Eq.~(\ref{appr}) that larger resonator decay rates will ultimately suppress the stabilization fidelity when $\kappa>2g\epsilon_2$. This fact can be understood when considering the system dynamics at the level of quantum trajectories: fast resonator decay leads to frequent jumps projecting the system state to a quantum state with definite photon number -- an effect similar to that of repeated projective measurements of the resonator's occupation number. For a large resonator decay rate, the coherent evolution between states $\vert 0,g\rangle$ and $\vert 1,e\rangle$ will thus be persistently interrupted, trapping the system in $\vert 0,g\rangle$ through the quantum Zeno effect. Therefore, exceedingly large resonator decay rates will ultimately slow down the increase of the magnitudes of population in state $\vert 1,e\rangle$ and $\vert 0,e\rangle$, which will lead to lower stabilization fidelities.

A combination of both red and blue sideband couplings in Eq.\ (\ref{Hrb}) enables the stabilization of mixed states centered on the $z$-axis of the Bloch sphere. As depicted in Fig.~{\ref{fig1}}(b), the interactions between states now become more complicated, since the three-level approximation is no longer appropriate. Different from blue-sideband coupling, red-sideband coupling promotes amplitude transfer between states $\vert m,e\rangle$ and $\vert m+1,g\rangle$ and may thus allow the system to access states with more than one photon inside the resonator. 

The particular qubit mixed state which is stabilized now depends on the magnitudes and relative phases of the red- and blue-sideband couplings. We fully characterize this mixed state by computing the ensemble averages $\langle \sigma_{x,y,z}\rangle$, and discuss their dependence on the couplings strengths.
The ensemble average of $\sigma_z$ in the non-equilibrium steady state is shown in Fig.~{\ref{fig2}}(b) as a function of the modulation strengths $g\epsilon_1$ and $g\epsilon_2$, using a fixed resonator decay rate. On average, the qubit acquires a larger portion of the excited state $\vert e\rangle$ for increasing $\epsilon_2/\epsilon_1$, and a larger portion of the ground state $\vert g\rangle$ for decreasing $\epsilon_2/\epsilon_1$. We note that the plot is approximately symmetric under exchange of $g\epsilon_1$ and $g\epsilon_2$ and, simultaneously, transforming $\sigma_z$ to $-\sigma_z$. Indeed, if we momentarily neglect the slow qubit dissipation, then the Lindblad master equation becomes invariant under interchange of $\sigma^-$ with $\sigma^+$, and $\epsilon_1$ with $\epsilon_2$. The qubit will be stabilized into a mixed state with equal weights of $\vert e\rangle$ and $\vert g\rangle$ with $\langle \sigma_z\rangle\approx 0$, when $g\epsilon_1$ equals $g\epsilon_2$. This symmetry breaks down when sideband coupling strengths become so small that qubit dissipation rates and the spurious ac-Stark shift cannot be neglected anymore.

Our numerical simulations show that ensemble averages of $\sigma_x$ and $\sigma_y$ vanish in the steady state. This can be understood as follows. Based on Fig.~\ref{fig1}(b), we can divide the system into two groups of states,
\begin{align*}
\mathrm{1.}\quad \vert 0,g\rangle,\,\vert 1,e\rangle,\,\vert 2,g\rangle,\,\vert 3,e\rangle \cdots\\
\mathrm{2.}\quad \vert 0,e\rangle,\,\vert 1,g\rangle,\,\vert 2,e\rangle,\,\vert 3,g\rangle \cdots
\end{align*}
The generation of coherent qubit superposition states of $\vert e\rangle$ and $\vert g\rangle$ would require hybridization of system states $\vert m,e\rangle$ and $\vert m,g\rangle$ with the same resonator occupation number $m$. However, red- and blue-sideband couplings can only hybridize states within each of the two groups, which excludes superpositions of $\vert m,e\rangle$ and $\vert m,g\rangle$. (Even if the initial state should present a nonzero matrix element $\langle m,e\vert \rho\vert m,g\rangle$, decoherence processes will effectively erase any such coherence.)

\subsection{Stabilizing the qubit along an arbitrary axis}



So far, we have discussed stabilization of the qubit in states along the $z$-axis of the Bloch sphere. This scheme has a natural generalization to qubit stabilization along an arbitrary axis  through the Bloch sphere. 

We employ the convention that the qubit excited state $\vert e\rangle$ resides at the north pole of the Bloch sphere. The axis specified by the unit vector $\mathbf{\hat{n}}$ has polar and azimuthal angles $\theta$ and $\phi$, respectively, with the pure qubit states $\vert \pm\mathbf{\hat{n}}\rangle$ located at the two points where the axis intercepts the Bloch sphere. Explicitly, the two pure states are given by
\begin{align}
\vert\mathbf{\hat{n}}\rangle&=\sin\frac{\theta}{2}\vert g\rangle+\mathrm{e}^{-i\phi}\cos\frac{\theta}{2}\vert e\rangle,\nonumber\\
\vert\mathbf{-\hat{n}}\rangle&=-\mathrm{e}^{-i\phi}\sin\frac{\theta}{2}\vert e\rangle+\cos\frac{\theta}{2}\vert g\rangle.
\end{align}
Points along the axis $\mathbf{\hat{n}}$ in the interior of the Bloch sphere represent mixed states composed of $\vert \mathbf{\hat{n}}\rangle$ and $\vert \mathbf{-\hat{n}}\rangle$, as usual.

We start by presenting how to stabilize the qubit in the pure state $\vert \mathbf{\hat{n}}\rangle$ on the Bloch sphere. Inspired by Fig.~\ref{fig1}(b), we aim for a Hamiltonian of the form
\begin{align}
H_{{\mathbf{\hat{n}}}B}=g\epsilon(a^\dagger\sigma^+_{\mathbf{\hat{n}}}+a\sigma^-_{\mathbf{\hat{n}}}),
\label{Heff}
\end{align}
analogous to Eq.~(\ref{Hb}). Here, $\sigma_{\mathbf{\hat{n}}}^{\pm}$ are defined via $\sigma_\mathbf{\hat{n}}^{\pm}\vert \mathbf{\mp\hat{n}}\rangle=\vert{\mathbf{\pm\hat{n}}}\rangle$. 
For the special case of $\theta=0$, this Hamiltonian reduces to the blue-sideband coupling. We call this Hamiltonian an effective blue-sideband coupling for state $\vert \mathbf{\hat{n}}\rangle$, which couples the system states $\vert m+1,\mathbf{\hat{n}}\rangle$ to $\vert m,\mathbf{-\hat{n}}\rangle$. 


As before, we require the resonator decay rate $\kappa$ and the coupling strength $g\epsilon$ to be much greater than the qubit dissipation rates. As shown in Fig.~\ref{arbstabschem}, the effective blue-sideband coupling for axis $\mathbf{\hat{n}}$ opens up a decay channel from $\vert 0,-\mathbf{\hat{n}}\rangle$ to $\vert 0,\mathbf{\hat{n}}\rangle$ via hybridization of $\vert 0,-\mathbf{\hat{n}}\rangle$ and $\vert 1,\mathbf{\hat{n}}\rangle$ and fast resonator decay from $\vert 1,\mathbf{\hat{n}}\rangle$ to $\vert 0,\mathbf{\hat{n}}\rangle$. Relative to these fast dynamics, qubit relaxation and dephasing is slow, leading to infrequent transitions between the states $\vert m,\mathbf{\hat{n}}\rangle$ and $\vert m,-\mathbf{\hat{n}}\rangle$. The resulting effective rates are given by \cite{Yao}
\begin{align}
\label{gamma-}
\tilde{\gamma}^-=&\;\gamma\cos^4\frac{\theta}{2}+\frac{\Gamma_{\varphi}}{2}\sin^2\theta,\nonumber\\
\tilde{\gamma}^+=&\;\gamma\sin^4\frac{\theta}{2}+\frac{\Gamma_{\varphi}}{2}\sin^2{\theta},\nonumber\\
\tilde{\Gamma}_{\varphi}=&\;\frac{\gamma}{2}\sin^2\theta+\Gamma_\varphi \cos^2\theta,
\end{align}
where $\tilde{\gamma}^\mp$ are the transition rates from qubit state $\vert \mathbf{\hat{n}}\rangle$ to $\vert \mathbf{-\hat{n}}\rangle$ (and reverse), and $\tilde{\Gamma}_\varphi$ is the effective dephasing rate. Since all three are much smaller than the resonator decay rate $\kappa$ and sideband coupling strength $g\epsilon$, the effective decay from $\vert 0,-\mathbf{\hat{n}}\rangle$ to $\vert 0,\mathbf{\hat{n}}\rangle$ dominates the dynamics and thus stabilizes the qubit in the state $\vert \mathbf{\hat{n}}\rangle$. 

\begin{figure}
\includegraphics[width=6.5cm]{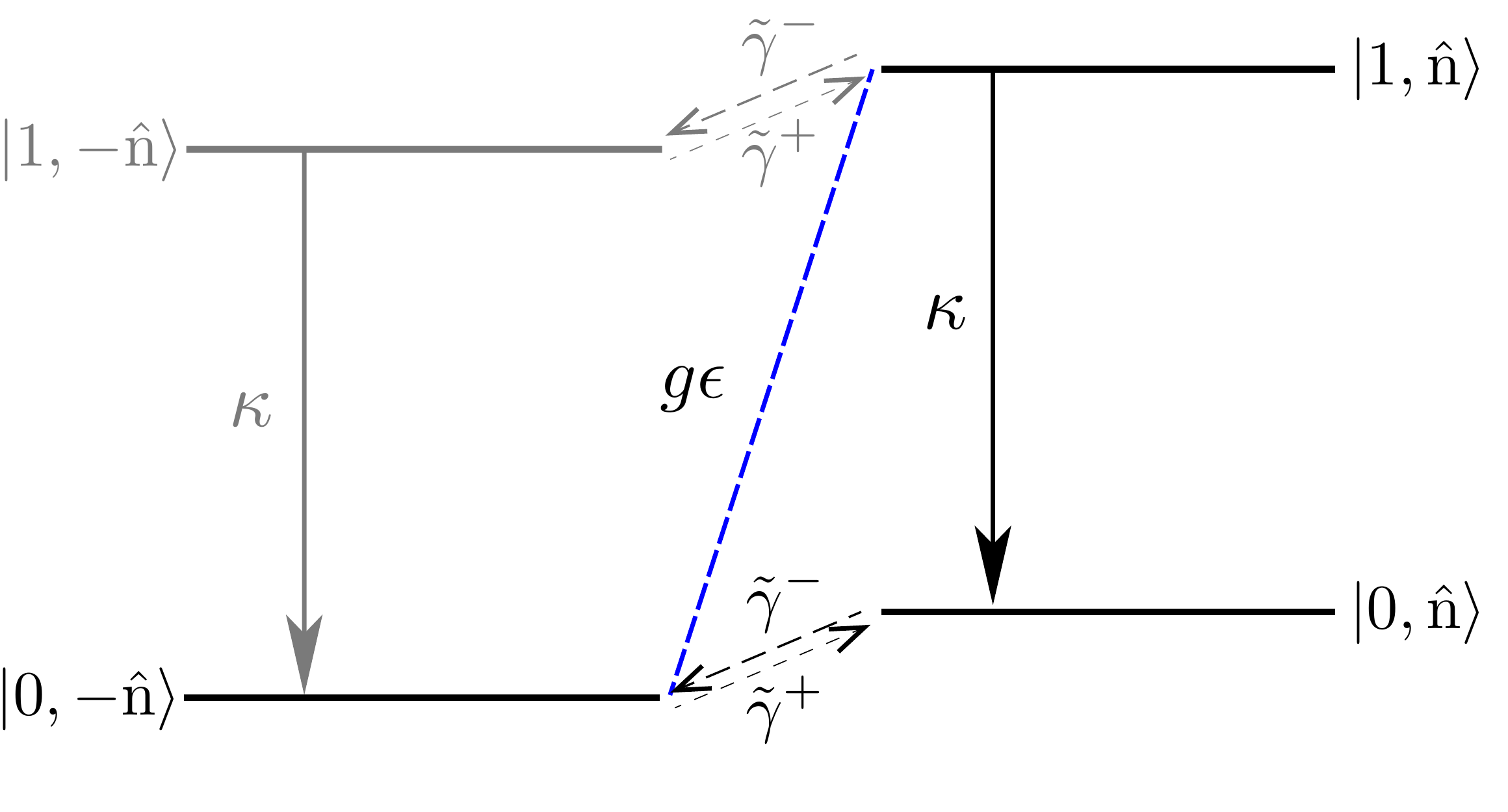}
\caption{Diagram of the arbitrary-axis stabilization scheme. With $\kappa\gg\tilde{\gamma}^-, \tilde{\gamma}^+$, the population in $\vert 1,-\mathbf{\hat{n}}\rangle$ can be safely neglected and the system is stabilized in $\vert 0,\mathbf{\hat{n}}\rangle$.}
\label{arbstabschem}
\end{figure}

We next show how to generate the desired Hamiltonian in Eq.~(\ref{Heff}) with our circuit-QED device. We first expand $\sigma_\mathbf{\hat{n}}^{\pm}$ in the Pauli matrix basis as
\begin{align}
\label{snpm}
&\sigma^\pm_{\mathbf{\hat{n}}}=\exp(-i\frac{\phi}{2}\sigma_z)\exp(-i\frac{\theta}{2}\sigma_y)\sigma^\pm\exp(i\frac{\theta}{2}\sigma_y)\exp(i\frac{\phi}{2}\sigma_z)\nonumber\\
&=\frac{1}{2}\sigma^\pm(\cos\theta+1)\mathrm{e}^{\mp i\phi}+\frac{1}{2}\sigma^\mp(\cos\theta-1)\mathrm{e}^{\pm i\phi}-\frac{1}{2}\sin\theta\sigma_z.
\end{align}
For simplicity (and without loss of generality) we set the azimuthal angle $\phi=0$ and defer the discussion of nonzero $\phi$ to the subsequent subsection. This way, we can plug the expression of $\sigma^\pm_{\mathbf{\hat{n}}}$ into Eq.~(\ref{Heff}) to obtain
\begin{align}
H_{\mathbf{\hat{n}}B}=&\;\frac{1}{2}g\epsilon(\cos\theta-1)(a^\dagger \sigma^-+a\sigma^+)\nonumber\\
&+\frac{1}{2}g\epsilon(\cos\theta+1)(a^\dagger\sigma^++a\sigma^-)\nonumber\\
&-\frac{1}{2}g\epsilon\sin\theta(a^\dagger+a)\sigma_z.
\label{Heff2}
\end{align}
Here, $H_{\mathbf{\hat{n}}B}$ denotes the effective blue-sideband coupling for state $\vert \mathbf{\hat{n}}\rangle$. This Hamiltonian is a combination of the red- and blue-sideband couplings, as well as a longitudinal coupling between the qubit and the resonator \cite{Longitude}. The latter can be generated by switching on a Rabi drive,
\begin{align}
\label{Rabi_driving}
H_{d}&=\xi(\sigma^-\mathrm{e}^{i\omega_3 t}+\sigma^+\mathrm{e}^{-i\omega_3 t}),
\end{align}
driving the qubit at the resonator frequency $\omega_3=\omega_r$ with strength $\xi$. This drive gives rise to a longitudinal coupling of the form
\begin{align}
H_d^\prime=-g'\xi(a^\dagger+a)\sigma_z,
\label{longitudinal}
\end{align}written in the dressed basis of the appropriate rotating frame. (We have dropped several fast-oscillating terms here.) The Rabi drive is realized by slightly modulating the fluxonium's penetrating flux. The details can be found in Appendix A. Therefore, the Hamiltonian in Eq.~(\ref{Heff2}) can be generated by tuning the strengths of the red- and blue-sideband couplings as well as the Rabi drive to match
\begin{align}
g\epsilon_1=&\;\frac{1}{2}g\epsilon(\cos\theta-1),\nonumber\\
g\epsilon_2=&\;\frac{1}{2}g\epsilon(\cos\theta+1),\nonumber\\
g^\prime\xi=&\;\frac{1}{2}g\epsilon\sin\theta,
\label{parameterchoice}
\end{align}
respectively.
\begin{figure}
\includegraphics[width=9cm]{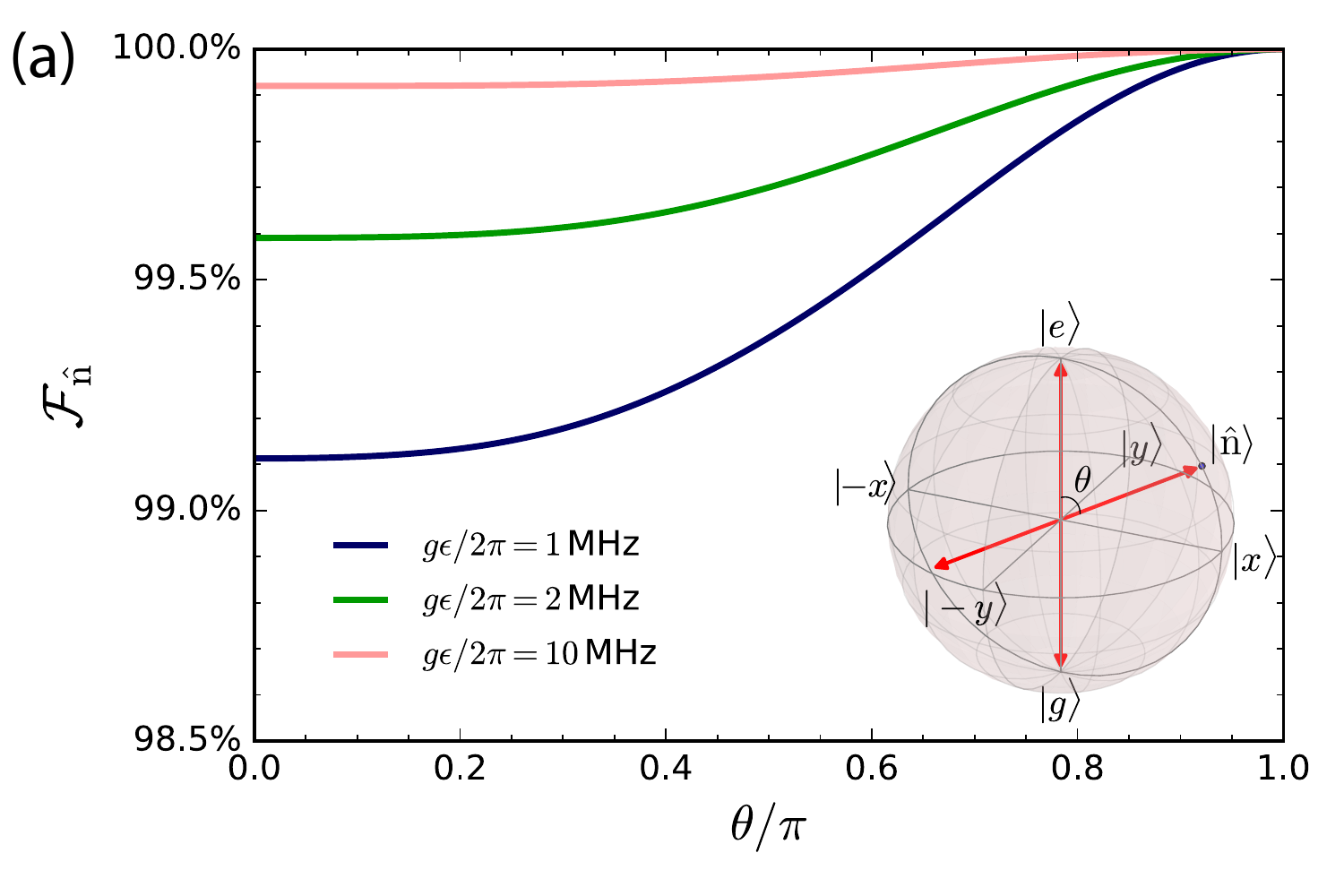}
\includegraphics[width=9cm]{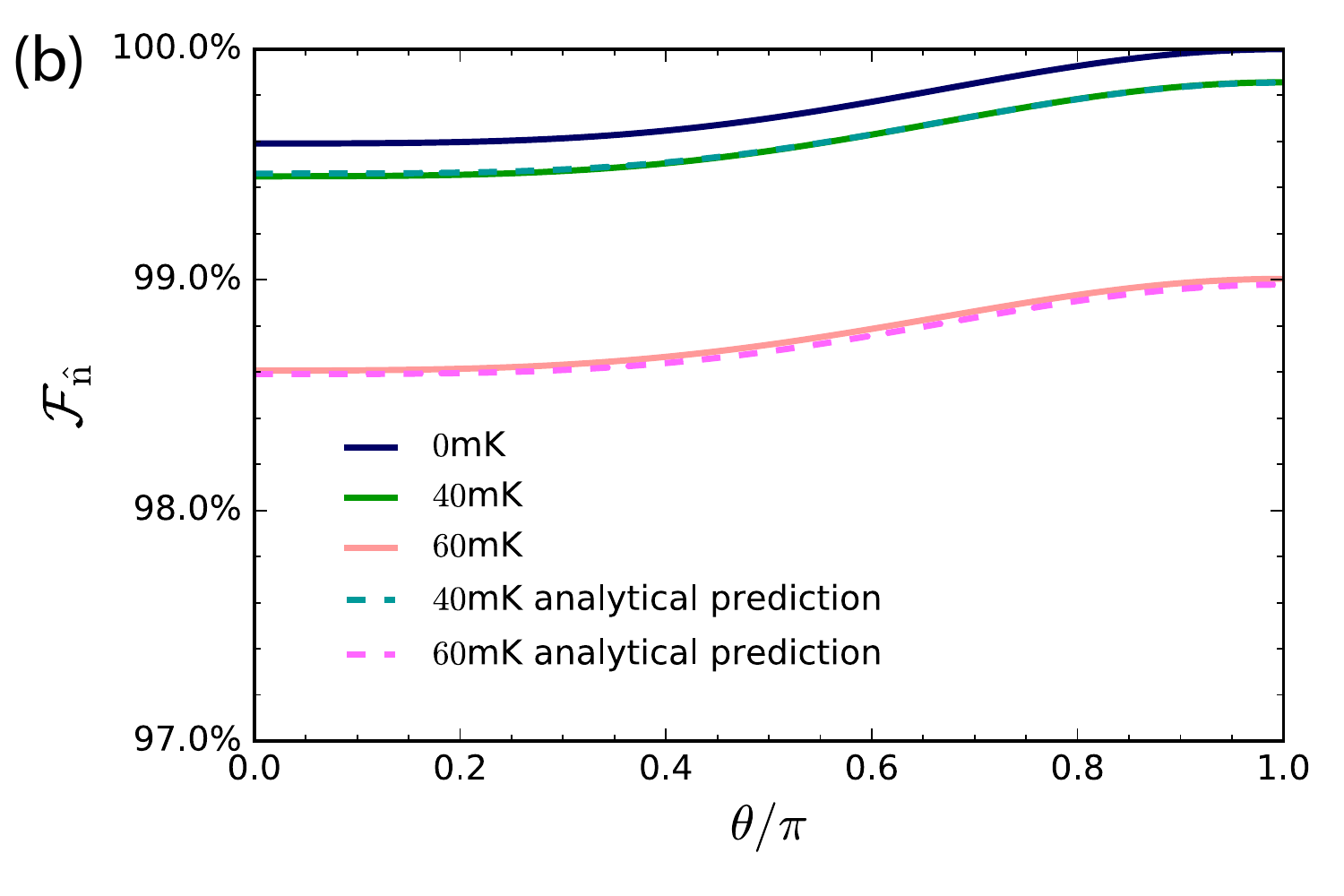}
\caption{(a) Stabilization fidelity for states along an axis $\mathbf{\hat{n}}$ in the $x$-$z$ plane, as a function of the polar angle $\theta$ (see in subplot). The three curves depict results for different strengths of the effective blue-sideband coupling. (Temperature and resonator decay rate are chosen as $T=15\,$mK, $\kappa/2\pi=4\,$MHz.) (b) Dependence of the fidelity on temperature, using $\kappa/2\pi=2g\epsilon/2\pi=4\,$MHz. Dashed curves show the analytical prediction from Eq.~(\ref{thermalcorrection}). In both graphs, we choose $\gamma=0.1\,$MHz, $\Gamma_{\varphi}=0.1\,$MHz and $\chi/2\pi=0.5\,$MHz. Excitation energies for the resonator and qubit are set to $4.89\,$GHz and $5.99\,$GHz, respectively. }\label{Fidelity}
\end{figure}

For a sideband coupling strength of $g\epsilon/2\pi = 2\,$MHz and resonator decay rate of $\kappa=2g\epsilon$, we can obtain state fidelities for $\vert \mathbf{\hat{n}}\rangle$ of up to 99.5\%, see Fig.\ref{Fidelity}(a). At zero temperature, the stabilization fidelity can also be obtained analytically based on the three-level model, and is approximately
\begin{align}
\label{F_n}
\mathcal{F}_\mathbf{\hat{n}}=\sqrt{\langle \mathbf{\hat{n}}\vert \rho_q\vert \mathbf{\hat{n}}\rangle} \approx \sqrt{1-\left[\frac{2g\epsilon}{\kappa}+\frac{\kappa}{2g\epsilon}\right]\frac{\tilde{\gamma}^-}{2g\epsilon}},
\end{align}
Details of the derivation are given in Appendix C.

Within the same approximation, we can further predict the stabilization fidelity at finite temperatures, and confirm that our scheme is robust to realistic levels of thermal excitations. The approximate relation between the stabilization fidelity and temperature is given by
\begin{align}
\mathcal{F}_\mathbf{\hat{n}}(T)\approx&\sqrt{\mathcal{F}_\mathbf{\hat{n}}^{2}(0)-\exp(-\hbar\omega_r/k_BT)\rho_{22}^{(0)}}.
\label{thermalcorrection}
\end{align}
where $\mathcal{F}_\mathbf{\hat{n}}(T)$ denotes the state fidelity of $\vert \mathbf{\hat{n}}\rangle$ obtained at temperature $T$. The quantity, $\rho_{22}^{(0)}$, represents the occupation probability for state $\vert 0,\mathbf{\hat{n}}\rangle$ at zero temperature, and is very close to 1 in our scheme; see again Appendix C for further details. The above expression shows that, to leading order, the influence of finite temperatures is directly determined by the comparison between resonator excitation energy $\hbar\omega_r$ and thermal excitation energy $k_BT$. We can thus suppress the influence of temperature by using a resonator with sufficiently large frequency while preserving the parameters of the qubit. Results shown in Fig.~\ref{Fidelity}(b) confirm that our scheme is robust with respect to thermal fluctuations at realistic operating temperatures and practical circuit parameters.


One can, in addition, generate an effective red-sideband coupling for $\vert \mathbf{\hat{n}}\rangle$, defined as $H_{\mathbf{\hat{n}}R}=g\epsilon(a^\dagger \sigma^-_{\mathbf{\hat{n}}}+a \sigma^+_{\mathbf{\hat{n}}})$. (Note that with $\sigma^\pm_{\mathbf{\hat{n}}}=\sigma^\mp_{-\mathbf{\hat{n}}}$, we have $H_{\mathbf{\hat{n}}R}=H_{-\mathbf{\hat{n}}B}$.) A combination of $H_{\mathbf{\hat{n}}B}$ and $H_{\mathbf{\hat{n}}R}$ can then stabilize the qubit in a mixed state of $\vert \mathbf{\hat{n}}\rangle$ and $\vert -\mathbf{\hat{n}}\rangle$, similar to our previous discussion and results in Figs.~\ref{fig1}(b) and \ref{fig2}(b). In other words, we can stabilize the qubit in a state corresponding to an arbitrary point along the axis defined by $\mathbf{\hat{n}}$. In the next subsection, we will discuss how to tune the state's azimuthal angle $\phi$, so that we can freely manipulate the axis $\mathbf{\hat{n}}$, and effectively stabilize the qubit for any point on and inside the Bloch sphere, at will.

\subsection{Azimuthal angle and phase matching}
So far, we have set the phases of the modulation and drive tones to zero at $t=0$, see Eqs.~(\ref{gt}) and (\ref{Rabi_driving}). This special choice only enables stabilization in the $\phi=0$ plane. To generalize this and stabilize states with arbitrary azimuthal angle $\phi$, detailed control of the phases is needed. We shall denote the phases of the three tones at time $t$ by
\begin{align}
P_n=\omega_n t+\nu_n,
\label{phases}
\end{align}
where $n=1,2$ stand for red- and blue-sideband modulation tones, and $n=3$ for the Rabi drive tone. For the latter, we set $\nu_3=0$ without loss of generality. The choice of the three frequencies yields the relations $\omega_1+\omega_2=2\omega_r=2\omega_3$ and $\omega_2-\omega_1=2\omega_q$. In the dressed bases of the appropriately rotating frame, the effective Hamiltonian in the presence of all three drives is then given by
\begin{align}
H=&\;g\epsilon_1(a^\dagger\sigma^-\mathrm{e}^{-i\nu_1}+\mathrm{H.c.})+g\epsilon_2(a^\dagger \sigma^+\mathrm{e}^{-i\nu_2}+\mathrm{H.c.})\nonumber\\
&-\frac{g'\xi}{\Delta}(a^\dagger+a)\sigma_z.
\label{Hrbd2}
\end{align}
Calculation shows that by tuning the strengths and phases of the three tones the Hamiltonian in Eq.~(\ref{Hrbd2}) can indeed generate the effective blue-sideband Hamiltonian
\begin{align}
H_{\mathbf{\hat{n}}B}=g\epsilon(a^\dagger \sigma^+_{\mathbf{\hat{n}}}+\mathrm{H.c.}),
\label{Hnbdelta}
\end{align}
if the drive strengths and phases satisfy the following conditions. First, the three phases from Eq.~(\ref{phases}) must obey
\begin{align}
P_1+P_2-2P_3
=\nu_1+\nu_2
=0.
\label{phasematching}
\end{align}
This relation reduces to one among the initial phases due to the frequency match among the three tones, i.e., $\omega_1+\omega_2=2\omega_3$.
Second, for the azimuthal angle $\phi$, we require
\begin{align}
(P_2-P_1)/2-\omega_q t
=(\nu_2-\nu_1)/2
=\phi.
\label{azimuthal}
\end{align}
Since $\omega_2-\omega_1=2\omega_q$, the azimuthal angle is simply determined by the initial phases of the modulation tones, $\nu_1$ and $\nu_2$. Third, the strengths of the three tones must meet the conditions of Eq.~(\ref{parameterchoice}) to set the desired polar angle $\theta$. 


To access arbitrary azimuthal angles, we thus require frequency matching and phase stability. Appendix B shows one technique that can generate the three tones based on two independent tones, through which Eq.~(\ref{phasematching}) is automatically satisfied.

\section{Fast stabilization and critical damping}
The time needed for stabilizing the qubit in a desired pure state is crucial for applications such as fast qubit initialization and reset. The time scale for pure-state stabilization is mainly set by $g\epsilon$ and $\kappa$. To make this statement more quantitative, we follow the dynamics of the axis $\mathbf{\hat{n}}$ stabilization scheme as described by the Lindblad master equation. Neglecting the population amplitude associated with $\vert 1,-\mathbf{\hat{n}}\rangle$, the stabilization process can be approximately described by the following set of differential equations:
\begin{align}
\frac{\mathrm{d}\rho_{33}}{\mathrm{d}t}=&\;2g\epsilon C-\kappa \rho_{33},\nonumber\\\label{differentialequation}
\frac{\mathrm{d}\rho_{11}}{\mathrm{d}t}=&\;-2g\epsilon C,\\\nonumber
\frac{\mathrm{d}C}{\mathrm{d}t}=&\;g\epsilon(\rho_{11}-\rho_{33})-\frac{1}{2}\kappa C,
\end{align}
see Appendix C for the detailed derivation. In the expression above, $\rho_{11}$ and $\rho_{33}$ are the occupation probabilities for $\vert 0,-\mathbf{\hat{n}}\rangle$ and $\vert 1,\mathbf{\hat{n}}\rangle$, respectively. The quantity $C$ denotes the imaginary part of the off-diagonal density matrix element for states $\vert 0,-\mathbf{\hat{n}}\rangle$ and $\vert 1,\mathbf{\hat{n}}\rangle$, i.e., $C=\mathrm{Im}[\langle 0,-\mathbf{\hat{n}}\vert \rho\vert 1,\mathbf{\hat{n}}\rangle]$.

These three first-order differential equations can be turned into a third-order differential equation for $\rho_{11}$,
\begin{align}
\frac{\mathrm{d}^3\rho_{11}}{\mathrm{d}t^3}+\frac{3}{2}\kappa\frac{\mathrm{d}^2\rho_{11}}{\mathrm{d}t^2}+(4g^2\epsilon^2+\frac{1}{2}\kappa^2)\frac{\mathrm{d}\rho_{11}}{\mathrm{d}t}+2\kappa g^2\epsilon^2 \rho_{11}=0,
\label{differentialequation2}
\end{align}
with an associated characteristic equation
\begin{align}
(\lambda+\frac{1}{2}\kappa)(\lambda^2+\kappa\lambda+4g^2\epsilon^2)=0.
\label{characteristic}
\end{align}
Similar to the classical damped harmonic oscillator, the stabilization process can be under-damped, critically-damped, or over-damped, depending on the nature of the roots of Eq.~(\ref{characteristic}).  Critically-damped stabilization occurs for $\kappa=4g\epsilon$, at which point all three roots of Eq.~(\ref{characteristic}) become real. Resonator dissipation rates deviating from this working point lead to under-damped or over-damped stabilization instead. For a fixed resonator dissipation rate, different sideband coupling strengths can also lead to all three damping types. 

Fig.~{\ref{Stabilizationprocess} shows the stabilization processes for different coupling strengths $g\epsilon$ at fixed $\kappa$, for stabilizing the qubit in its excited state $\vert e\rangle$ and in the superposition $\vert x\rangle=\frac{1}{\sqrt{2}}(\vert e\rangle+\vert g\rangle)$. As $g\epsilon$ is decreased, we find behavior characteristic of the three damping types. Compared with critically-damped stabilization, a slightly under-damped case may help the system reach the steady state faster, since the tiny oscillations, arising from complex roots of Eq.~(\ref{characteristic}), are almost negligible as evidenced by numerical simulations. For our chosen system parameters, we find that $g\epsilon\approx \kappa/2.6$ yields the quickest stabilization.
\begin{figure}
\includegraphics[width=9cm]{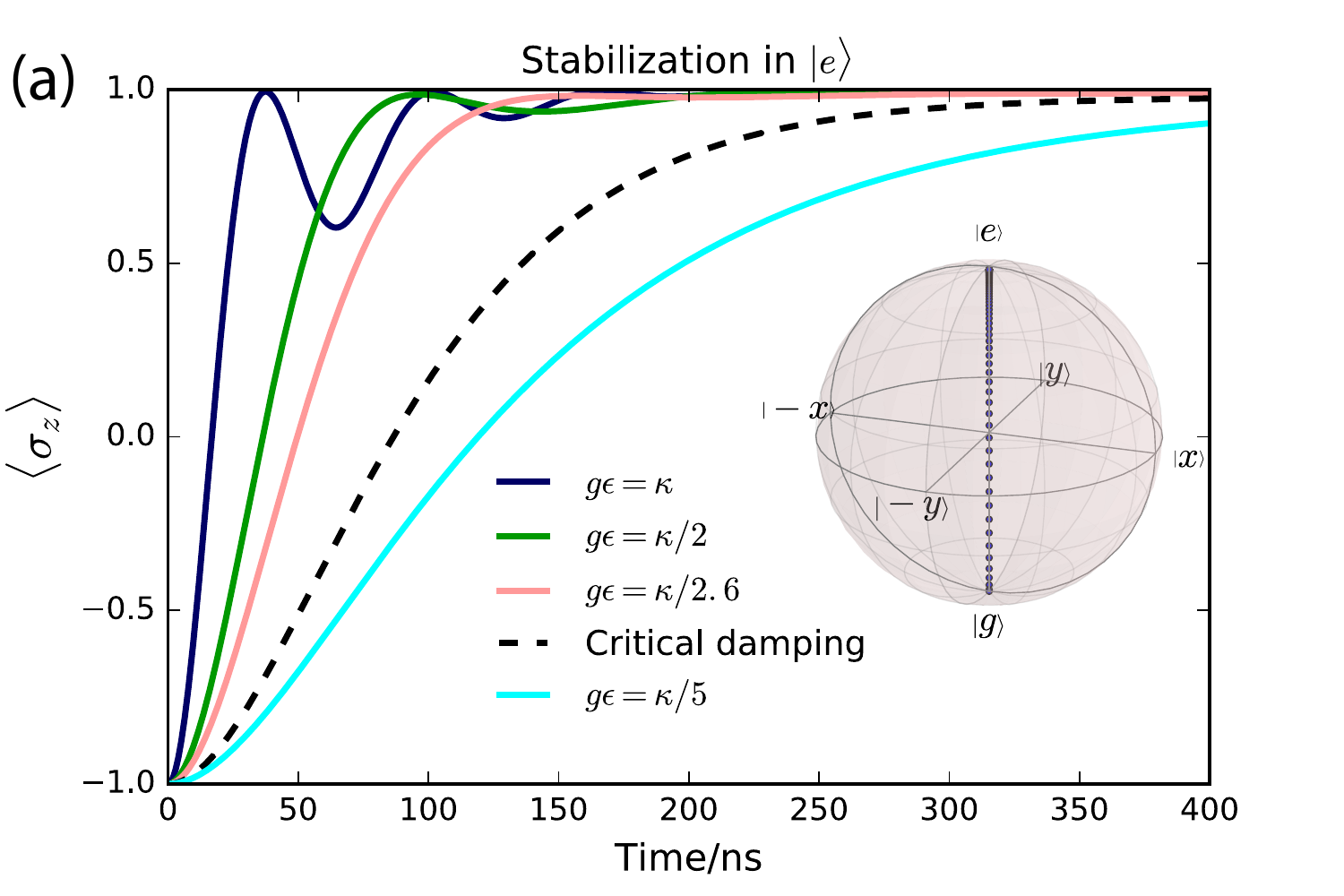}
\includegraphics[width=9cm]{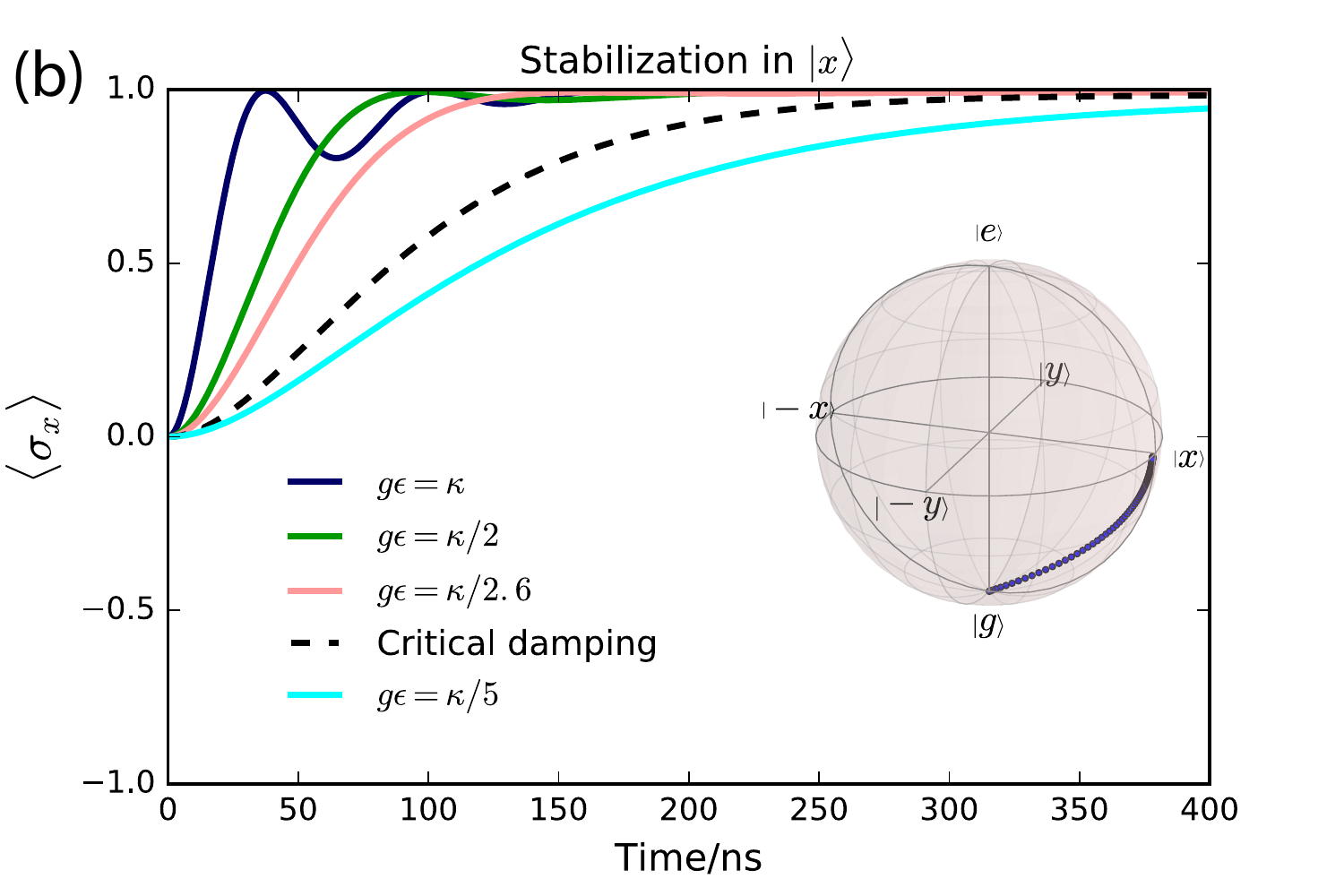}
\caption{Stabilization processes in time domain. The qubit is initialized in the ground state. Shown is the expectation of $\langle \sigma_{\mathbf{\hat{n}}}\rangle$ when targeting (a) the excited state and (b) state $\vert x\rangle$. The insets show the stabilization dynamics in terms of $\rho(t)$. ($\kappa/2\pi$ is set to $8\,$MHz,  $T=15\,$mK, $\gamma=0.1\,$MHz, $\Gamma_{\varphi}=0.1\,$MHz and $\chi/2\pi= 0.5\,$MHz.)}
\label{Stabilizationprocess}
\end{figure}

The stabilization time is set by $2/\kappa$ which is the characteristic time for the critically-damped stabilization process. With realistic parameters, as chosen for Fig.~{\ref{Stabilizationprocess}, the stabilization can be completed within around $100\,$ns. 
\section{Conclusion}
In conclusion, we have presented and analyzed the performance of a universal single-qubit stabilization scheme. By modulating the external flux penetrating the coupler, red- and blue-sideband couplings are generated between the qubit and resonator. The combined use of both couplings and a Rabi drive enables the generation of a special coupling between the qubit and lossy resonator, as in Eq.~(\ref{Heff}). With it, the qubit can be autonomously cooled towards any point on the Bloch sphere, with fidelities over 99.5\%. Such stabilization can be completed within around 100$\,$ns for practical parameters. Stabilizing the qubit in mixed states, i.e., points inside the Bloch sphere, is possible by tuning the strengths and phases of modulation and Rabi drive tones. Our scheme is robust with respect to realistic temperature fluctuations. 

For the stabilization of pure qubit states, we have shown that the system dynamics can be captured by a three-level model, and can be analytically solved both for the steady state and dynamical stabilization process. The dynamical process can be understood by a third-order differential equation, allowing us to distinguish between under-damped, critically-damped and over-damped stabilization. The idea of three types of stabilization processes and optimization of stabilization time might also be explored in other schemes, and future generalizations to multi-qubit states.

\section{Acknowledgements}
We thank Peter Groszkowski and Andy C. Y. Li for their valuable comments and discussions. This research was supported by the Army Research Office through Grants No.\,W911NF-15-1-0421 and No.\,W911NF-18-1-0125 and by the National Science Foundation under Grant No.\,PHY-1653820.
\appendix
\section{Circuit analysis}
The circuit of the considered device is shown in Fig.~{\ref{fig0}} and yields the Lagrangian
\begin{align}
&L=\frac{1}{2}C_r\dot{\Phi}_a^2+\frac{1}{2}C_{J1}\dot{\Phi}_b^2+\frac{1}{2}C_s(\dot{\Phi}_b-\dot{\Phi}_c)^2+\frac{1}{2}C_q\dot{\Phi}_c^2\nonumber\\
&+\frac{1}{2}C_{qr}(\dot{\Phi}_a-\dot{\Phi}_b)^2
-\frac{1}{2L_r}(\Phi_a-\Phi_b)^2-\frac{1}{2L_q}\Phi_c^2\nonumber\\
&+E_{J2}\cos{(2\pi\frac{\Phi_c+\Phi_{\mathrm{ext,fl}}}{\Phi_0}})+E_{J1\mathrm{eff}}(t)\cos{\frac{2\pi\Phi_b}{\Phi_0}}.
\label{lagrangian}
\end{align}
The SQUID loop's Josephson energy $E_{J1\mathrm{eff}}(t)=2E_{J1}\cos[\varphi_{\mathrm{sq}}(t)/2]$ is tuned by the external flux $\varphi_{\mathrm{sq}}(t)=2\pi\Phi_{\mathrm{ext}}(t)/\Phi_0$ which is modulated around its dc value using two modulation tones, i.e., $\varphi_{\mathrm{sq}}(t)=\overline{\varphi}_{\mathrm{sq}}-d_1\cos\omega_1t-d_2\cos\omega_2t$, with $d_1,d_2\ll 1$. 
As long as modulation amplitudes for the external flux remain small, we can expand $E_{J1\mathrm{eff}}(t)$ into its dc value and a small time-varying part,
\begin{align}
E_{J1\mathrm{eff}}(t)=E_{J1\mathrm{eff}}^{(0)}+E_{J1\mathrm{eff}}^\prime(t),
\end{align}
where $E_{J1\mathrm{eff}}^{(0)}$ is the time-average of $E_{J1\mathrm{eff}}(t)$. $E'_{J1\mathrm{eff}}(t)$ can be approximated as
\begin{align}
E'_{J1\mathrm{eff}}(t)\approx(2\epsilon_1\cos\omega_1t+2\epsilon_2\cos\omega_2t)E_{J1\mathrm{eff}}^{(0)},
\end{align}
where $2\epsilon_n\approx\sin(\varphi_{\mathrm{ext}}/2)d_n/2$ $(n=1,2)$. (In this definition of $\epsilon_n$, a factor of 2 is included for more convenient notation in the main text.)
The Hamiltonian in Eq.~(\ref{FullH}) can be obtained from Eq.\ \eqref{lagrangian} by a Legendre transformation. 

The coupler mode only serves a passive role by tuning the coupling between the resonator and qubit. For this purpose, we choose the energy scales of the relevant circuit parameters as listed in Table~\ref{label1}.
\begin{table}[h]
\caption{Energy scales of circuit parameters.}
\squeezetable
\begin{center}
\begin{tabular}{  m{9em}  m{6em} } 
\hline\hline
Parameters & Energy Scale   \\ 
\hline
$E^{(0)}_{J1\mathrm{eff}}/2\pi$ & $\sim 1000\,$GHz  \\ 
\hline
$E_{Lr}/2\pi$ & $\sim 50\,$GHz \\ 
\hline
$E_{J2}/2\pi$ & $\sim 10\,$GHz \\ 
\hline
$E_{c}/2\pi$ & $\sim 4\,$GHz \\ 
\hline
$E_{Lq}$, $E_{ac}$, $E_{bc}/2\pi$ & $\sim 300\,$MHz\\
\hline
$E_a$, $E_b/2\pi$ & $\sim 100\,$MHz\\
\hline\hline
\end{tabular}
\end{center}
\label{label1}
\end{table} 
By design, the Josephson energy $E^{(0)}_{J1\mathrm{eff}}$ is the largest energy scale so that the coupler mode $b$ has excitation energies far exceeding those of the qubit and resonator. The potential energy of mode $b$ is dominated by the term $-E^{(0)}_{J1\mathrm{eff}}\cos\varphi_b$ and, since $E^{(0)}_{J1\mathrm{eff}}\gg E_b$, low-lying wave functions will be localized around $\varphi_b=0$. The corresponding oscillator length is given by $(8E_b/E_{J1\mathrm{eff}}^{(0)})^{1/4}\ll 1$. We approximate the Hamiltonian by a second-order expansion in $\varphi_b$ which gives
\begin{align}
\label{GroupedH}
H=&\left[4E_bn_b^2+\frac{1}{2}\left(E_{Lr}+E_{J1\mathrm{eff}}(t)\right)\varphi_b^2\right]\nonumber\\
&+\left[4E_cn_c^2-E_{J2}\cos(\varphi_c+\varphi_{\mathrm{fl}})+\frac{E_{Lq}}{2}\varphi^2_c\right]\nonumber\\
&+\left[4E_an_a^2+\frac{E_{Lr}}{2}\varphi^2_a\right]+E_{ab}n_an_b+E_{bc}n_bn_c\nonumber\\
&-E_{Lr}\varphi_a\varphi_b.
\end{align}
In terms of annihilation and creation operators for the $a$ and $b$ modes as well as eigenstates $\{|j\rangle\}$ of the $c$ (qubit) mode, the Hamiltonian can be rewritten in the form
\begin{align}
\label{Habc}
 H\approx &\;\Omega_a a^\dagger a+\Omega_b b^\dagger b+\sum_{j}E_j\vert j\rangle\langle j\vert\nonumber\\
 &+i(a^\dagger-a)\sum_{j,k}(g_{a;jk}\vert j\rangle \langle k\vert +\text{H.c.})\nonumber\\
 &+i(b^\dagger-b)\sum_{j,k}(g_{b;jk}\vert j\rangle \langle k\vert +\text{H.c.})\nonumber\\
 &+\Omega_{ab}(a^\dagger+a)(b^\dagger +b)+\Omega_{\mathrm{mod}}(t)(b^\dagger +b)^2.
\end{align}
Here, $\Omega_a$ and $\Omega_b$ are the excitation energies of the resonator and coupler, and $E_j$ is the energy of the qubit eigenstate $|j\rangle$.
We design $\Omega_b$ to be the largest excitation energy among the three degrees of freedom, setting $\Omega_b\sim 2\pi\times 20\,\mathrm{GHz}$ and $\Omega_a, E_1-E_0 \sim 2\pi \times 5\,\mathrm{GHz}$. 
$\Omega_{ab}$ is the coupling strength between the resonator and coupler due to the $\varphi_a\varphi_b$  term in Eq.~(\ref{GroupedH}). The coupling strengths between the qubit and resonator ($g_{a;jk}$) or coupler ($g_{b;jk}$) are due to terms involving $n_an_b$ and $n_bn_c$ in Eq.~(\ref{GroupedH}). These coefficients are given by
\begin{align}
\Omega_{ab}&=E_L\left[\frac{2E_a}{E_L}\right]^{\frac{1}{4}}\left[\frac{2E_b}{E_L+E^{(0)}_{J1\mathrm{eff}}}\right]^{\frac{1}{4}},\nonumber\\
g_{a;jk}&=E_{ac}\langle j\vert n_c\vert k\rangle\left[\frac{E_{Lr}}{32E_a}\right]^{\frac{1}{4}},\nonumber\\
g_{b;jk}&=E_{bc}\langle j\vert n_c\vert k\rangle\left[\frac{E_{Lr}+E^{(0)}_{J1\mathrm{eff}}}{32E_b}\right]^{\frac{1}{4}}.
\label{fgh}
\end{align}
All of them are small quantities compared with the excitation energies of the three modes, and can be treated perturbatively. $\Omega_{\mathrm{mod}}(t)=\sqrt{2E_b/(E_L+E^{(0)}_{J1\mathrm{eff}})}E^\prime_{J1\mathrm{eff}}(t)$ denotes the strength of time-dependent modulation on the coupler mode. 

Since the coupler remains in its ground state, we may eliminate it adiabatically from the Hamiltonian.
To this end, we adopt a Bogoliubov transformation \cite{Bogoliubov} removing the static coupling term between  resonator and coupler. As a result of the transformation, the coefficients of the remaining terms in Eq.\ \eqref{Habc} will be shifted. Second, a Schrieffer-Wolff transformation \cite{SWtransformation,Guanyu} decoupling the qubit from the other two modes is applied.
Switching to the new dressed basis, all static couplings among the three modes are removed. The coupler's annihilation operator $b$ is transformed into
\begin{align}
b\;\rightarrow \;b-\frac{\Omega_{ab}}{\Delta_{ab}}a-\frac{\Omega_{ab}}{\Sigma_{ab}}a^{\dagger}+\sum_{j,k}\frac{\text{i}g_{b;kj}}{\Delta_{b;kj}}\vert j\rangle\langle k\vert,
\label{btransformation}
\end{align}
where $\Delta_{ab}=\Omega_b-\Omega_a$, $\Sigma_{ab}=\Omega_a+\Omega_b$, $\Delta_{b;kj}=\Omega_b-(E_k-E_j)$.  The time-dependent modulation term $\Omega_\mathrm{mod}(t)(b^\dagger+b)^2$ is transformed, to leading order, into
\begin{align*}
\Omega_{\mathrm{mod}}(t)\left[(b+b^\dagger)+\eta_a(a+a^\dagger)+\sum_{jk}(\eta_{jk}\vert j\rangle\langle k\vert+\text{H.c.})\right]^2,
\end{align*}
where $\eta_a\approx -2\Omega_b\Omega_{ab}/(\Omega_b^2-\Omega_a^2)$, and 
\begin{align*}
\eta_{jk}\approx \frac{\text{i}g_{b;kj}}{\Delta_{b;kj}} -\frac{\text{i}g_{b;jk}}{\Delta_{b;jk}}.
\end{align*}
With this, we finally obtain the effective Hamiltonian
\begin{align}
H=&\;\Omega_a a^\dagger a+\sum_jE_j\vert j\rangle\langle j\vert\nonumber\\
&+\sum_{j}\chi_{a,j}a^\dagger a\vert j\rangle\langle j\vert+\sum_j\kappa_j\vert j\rangle\langle j\vert\nonumber\\
&+\Omega_{\mathrm{mod}}(t)\left[\eta^2_a(a+a^\dagger)^2+\left(\sum_{jk}\eta_{jk}\vert j\rangle\langle k\vert+\text{H.c.}\right)^2\right]\nonumber\\
&+2\Omega_{\mathrm{mod}}(t)\eta_a (a^\dag+ a)\left(\sum_j\eta_{jk}\vert j\rangle\langle k\vert+\text{H.c.}\right),
\label{eq:effH}
\end{align}
describing the resonator and qubit modes only, where  $\chi_{a,j}$ and $\kappa_j$ stand for the dispersive shifts \cite{Guanyu}. When approximating the fluxonium qubit as a two-level system, we recover the Hamiltonian in Eq.~(\ref{FullH2}), with the coefficients given by
\begin{align}
\omega_r=&\;\Omega_a+\frac{\chi_{a,0}+\chi_{a,1}}{2},\nonumber\\
\omega_q=&\;(E_1+\kappa_1)-(E_0+\kappa_0),\nonumber\\
\chi=&\;\frac{\chi_{a,0}-\chi_{a,1}}{2}.
\end{align} In Eq.~(\ref{eq:effH}), the second to last line introduces small oscillations in the resonator and qubit energies, but can be neglected within the rotating-wave approximation. Terms in the last line of Eq.\ \eqref{eq:effH} give rise to time-dependent coupling between the resonator and qubit, and lead to the expression of $g(t)$ in Eq.~(\ref{gt}). The magnitude $g$ of that coupling is given by
\begin{align}
g=2\eta_a\eta_{01}E^{(0)}_{J1\mathrm{eff}}\sqrt{\frac{2E_b}{E_L+E^{(0)}_{J1\mathrm{eff}}}}.
\end{align}

Slight modulation of the fluxonium's reduced penetrating flux, $\varphi_{\mathrm{fl}}(t)=d_3\cos\omega_3t$, yields the Rabi drive in Eq.~(\ref{Rabi_driving}). To see this, we approximate
\begin{align*}
\cos(\varphi_c+\varphi_{\mathrm{fl}}(t))&\approx 
\cos\varphi_c-d_3\cos\omega_3t\sum_{jk}f_{jk}\vert j\rangle\langle k\vert,
\end{align*}
where $f_{jk}=\langle j\vert \sin\varphi_c\vert k\rangle$.
In the dressed basis, this drive gives terms involving $(a^\dagger +a)\vert 0\rangle\langle 0\vert$ and $(a^\dagger +a)\vert 1\rangle\langle 1\vert$, leading to the longitudinal coupling in Eq.~(\ref{longitudinal}). The coefficient $g'$ in Eq.~(\ref{longitudinal}) is given by
\begin{align}
g'=\frac{\alpha_0-\alpha_1}{2f_{01}},
\end{align} 
where
\begin{align}
\alpha_0&=\sum_j {f_{j0}}\left(-\frac{\text{i}g_{a;0j}}{\Delta_{a;0j}}\right)- {f_{0j}}\left(-\frac{\text{i}g_{a;j0}}{\Delta_{a;j0}}\right),\nonumber\\
\alpha_1&=\sum_j f_{j1} \left(-\frac{\text{i}g_{a;1j}}{\Delta_{a;1j}}\right)-f_{1j}\left(-\frac{\text{i}g_{a;j1}}{\Delta_{a;j1}}\right),
\end{align}
and $\Delta_{a;jk}=\Omega_a-(E_j-E_k)$. 

\section{Phase matching among three tones}
 In Section III.C, we noted that stabilization required phase matching, see Eqs.\ \eqref{phasematching} and \eqref{azimuthal}. We show here that the  three modulation and drive tones obeying the desired phase constraint can be generated by two tones.
 
We start with two coherent tones at the dressed resonator and qubit frequencies,
\begin{align}
H_1=\cos(\omega_q t+\upsilon_1),\qquad 
H_2=\cos(\omega_r t+\upsilon_2).
\label{newtones}
\end{align}
where we set amplitudes to 1, for simplicity. To generate the two modulation tones with frequencies $\omega_{1,2}$, we consider the product tone
\begin{align}\label{product}
H_1H_2&=\cos(\omega_q t+\upsilon_1)\cos(\omega_r t+\upsilon_2)\\
&=\frac{1}{2}\cos[(\omega_r-\omega_q)t+\upsilon_2-\upsilon_1]\nonumber\\
&\,+\frac{1}{2}\cos[(\omega_r+\omega_q)t+\upsilon_2+\upsilon_1]\nonumber\\
&=\frac{1}{2}\cos(\omega_1 t+\upsilon_2-\upsilon_1)+\frac{1}{2}\cos(\omega_2t+\upsilon_2+\upsilon_1)\nonumber
\end{align}
which is a superposition of tones $T_1$, $T_2$ with frequencies $\omega_1$ and $\omega_2$. Extracted via high- and low-pass filters
\begin{align*}
T_1=\cos(\omega_1 t+\upsilon_2-\upsilon_1),\quad
T_2=\cos(\omega_2t+\upsilon_2+\upsilon_1),
\end{align*}
can be used for the generation of red- and blue-sideband modulations. The Rabi drive tone can be directly generated from \begin{align*}
T_3=H_2=\cos(\omega_r t+\upsilon_2),
\end{align*} 
since $\omega_3=\omega_r$,  choosing $\upsilon_2=0$. One can confirm that the condition set in Eq.~(\ref{phasematching}) is automatically satisfied in this scheme. Moreover, the azimuthal angle, see Eq.\ \eqref{azimuthal}, is conveniently chosen by $\phi=\upsilon_1$.

\section{Analytical solution of three-level model}
We base our discussion of stabilization fidelity and time on a three-level model shown in Fig.~{\ref{arbstabschem}}. Specifically, we neglect residual population of state $\vert 1,-\mathbf{\hat{n}}\rangle$ and confine the dynamics of the system to a subspace spanned by $\vert 0,-\mathbf{\hat{n}}\rangle$, $\vert 0,\mathbf{\hat{n}}\rangle$ and $\vert 1,\mathbf{\hat{n}}\rangle$. In the case of arbitrary-state stabilization, the stabilization dynamics is governed by the Lindblad master equation
\begin{align}
\frac{\mathrm{d}\rho}{\mathrm{d}t}=&-i[H_{\mathrm{eff}},\rho]+\kappa\,\mathbb{D}[a]\rho\nonumber\\
&+\tilde{\gamma}^-\,\mathbb{D}[\sigma_{\mathbf{\hat{n}}}^-]\rho+\tilde{\gamma}^+\,\mathbb{D}[\sigma_{\mathbf{\hat{n}}}^+]\rho+\frac{\tilde{\Gamma}_\varphi}{2}\,\mathbb{D}[\sigma_{\mathbf{\hat{n}}}]\rho,
\label{masterseq}
\end{align}
where $H_{\mathrm{eff}}$ refers to Eq.~(\ref{Heff}), and $\sigma_{\mathbf{\hat{n}}}$ is defined as $\sigma_{\mathbf{\hat{n}}}=2\sigma_{\mathbf{\hat{n}}}^+\sigma_{\mathbf{\hat{n}}}^--1$. The decoherence rates $\tilde{\gamma}^-$, $\tilde{\gamma}^+$ and $\tilde{\Gamma}_\varphi$ were defined in Eq.~(\ref{gamma-}). The time evolution of the density matrix $\rho$ can be described in terms of four key components:
\begin{align}
\label{fulltimedependent}
\frac{\mathrm{d}\rho_{11}}{\mathrm{d}t}=&-2g\epsilon C+\tilde{\gamma}^-\rho_{22}-\tilde{\gamma}^+ \rho_{11}, \nonumber\\
\frac{\mathrm{d}\rho_{22}}{\mathrm{d}t}=&\;\kappa \rho_{33}-\tilde{\gamma}^-\rho_{22}+\tilde{\gamma}^+\rho_{11},\nonumber\\
\frac{\mathrm{d}\rho_{33}}{\mathrm{d}t}=&\;2g\epsilon C-\kappa\rho_{33},\nonumber\\
\frac{\mathrm{d}C}{\mathrm{d}t}=&\;g\epsilon(\rho_{11}-\rho_{33})-(\frac{1}{2}\kappa+\frac{1}{2}\tilde{\gamma}^++\tilde{\Gamma}_{\varphi})C.
\end{align}
Here, $\rho_{11}$, $\rho_{22}$ and $\rho_{33}$ give the probability amplitudes for the states $\vert 0,-\mathbf{\hat{n}}\rangle$, $\vert 0,\mathbf{\hat{n}}\rangle$ and $\vert 1,\mathbf{\hat{n}}\rangle$, and $C=\mathrm{Im}[\langle 0,-\mathbf{\hat{n}}\vert \rho\vert 1,\mathbf{\hat{n}}\rangle]$. Due to the constraint $\rho_{11}+\rho_{22}+\rho_{33}=1$, only three of these four equations are independent. Once qubit decoherence (with coefficients $\tilde{\gamma}^+$, $\tilde{\gamma}^-$ and $\tilde{\Gamma}_\varphi$) is neglected, we recover the differential equation (\ref{differentialequation}).

The stabilized state is obtained by setting all time derivatives in Eq.~(\ref{fulltimedependent}) to zero, and we obtain an exact expression for the stabilization fidelity:
\begin{align}
\mathcal{F}_\mathbf{\hat{n}}=\sqrt{1-\bigg[\frac{2g\epsilon}{\kappa}+\bigg(\frac{1}{2}\kappa+\frac{1}{2}\tilde{\gamma}^++\tilde{\Gamma}_{\varphi}\bigg)\bigg/g\epsilon\bigg]C},
\end{align}
where
\begin{align}
C=\frac{\tilde{\gamma}^-/(\tilde{\gamma}^++\tilde{\gamma}^-)}{\frac{2g\epsilon}{\kappa}+\frac{2g\epsilon}{\tilde{\gamma}^++\tilde{\gamma}^-}(1+\frac{\tilde{\gamma}^-}{\kappa})+(\frac{1}{2}\kappa+\frac{1}{2}\tilde{\gamma}^++\tilde{\Gamma}_{\varphi})\frac{1}{g\epsilon}}.
\end{align}
The approximate result for the stabilization fidelity of the qubit excited state $\vert e\rangle$, given in Eq.~(\ref{Fidelity}), is recovered by taking $\tilde{\gamma}^+=0$ and $\tilde{\gamma}^-=\gamma$.

Thermal fluctuations will generally lower the stabilization fidelity. The influence of temperature can be assessed by a perturbative treatment within the three-level model. For finite temperatures, we add the terms $\kappa_{\mathrm{th}}\,\mathbb{D}[a^\dagger]\rho$ and $\gamma_{\mathrm{th}}\,\mathbb{D}[\sigma^+]\rho$ to the Lindblad master equation \eqref{masterseq}, where $\kappa_{\mathrm{th}}=\kappa\exp(-\hbar\omega_r/k_BT)$ and $\gamma_{\mathrm{th}}=\gamma\exp(-\hbar\omega_q/k_BT)$. In the low-temperature limit ($\kappa_{\mathrm{th}}\ll \kappa$ and $\gamma_{\mathrm{th}}\ll\gamma$), we maintain $\langle 0,\mathbf{\hat{n}}\vert \rho\vert 0,\mathbf{\hat{n}}\rangle\approx 1$. Further, for $\omega_r\sim\omega_q$ and $\gamma\ll\kappa$, we also have $\gamma_{\mathrm{th}}\ll\kappa_{\mathrm{th}}$. As a result, we expect the leading corrections due to thermal excitations to be given by the excitation from $\vert 0,\mathbf{\hat{n}}\rangle$ to $\vert 1,\mathbf{\hat{n}}\rangle$ at  rate $\kappa_{\mathrm{th}}$. Within perturbation theory, the first-order corrections to our zero-temperature solutions $\rho^{(0)}_{ii}(i=1,2,3)$ and $C^{(0)}$ obey:
\begin{align}
0&=-2g\epsilon C^{(1)}+\tilde{\gamma}^-\rho^{(1)}_{22}-\tilde{\gamma}^+ \rho^{(1)}_{11},\nonumber\\
0&=\kappa_{\mathrm{th}}\rho^{(0)}_{22}+2g\epsilon C^{(1)}-\kappa\rho^{(1)}_{33},\nonumber\\
0&=g\epsilon(\rho^{(1)}_{11}-\rho^{(1)}_{33})-(\frac{1}{2}\kappa+\frac{1}{2}\tilde{\gamma}^++\Gamma_{\varphi})C^{(1)}.
\end{align}
With $g\epsilon$ and $\kappa$ far exceeding the qubit dissipation rates, we can infer from the first equation that $C^{(1)}$ should be much smaller than $\rho^{(1)}_{11}$ and $\rho^{(1)}_{22}$, and thus can be neglected in the second and third equation. As a result, we find the relation
\begin{align}
\rho_{11}^{(1)}\approx\rho^{(1)}_{33}\approx \frac{\kappa_{\mathrm{th}}}{\kappa}\rho^{(0)}_{22},
\label{fidelity_th}
\end{align} 
shown in Eq.~(\ref{thermalcorrection}).
\bibliography{mybib}
\end{document}